\documentclass[galaxies,review,accept,oneauthor,pdftex,12pt,a4paper]{mdpi} 
\newcommand{\cc}[2]{c{#1\atopwithdelims()#2}}
\newcommand{\oo}[2]{\left(#1\left|#2\right.\right)}
\newcommand{\Th}[2]{\vt\left[^{#1}_{#2}\right]}
\newcommand{\Thbar}[2]{\bar\vt\left[^{#1}_{#2}\right]}

\newcommand{\vt}{\vartheta}
\newcommand{\non}{\nonumber}

\def\beq{\begin{equation}}
\def\eeq{\end{equation}}
\def\beqn{\begin{eqnarray}}
\def\eeqn{\end{eqnarray}}

\def\AEF{Faraggi, A.E.}

\def\JCAP#1#2#3{{\em JCAP}              {\bf #2}, {\em  #1},  #3}
\def\JHEP#1#2#3{{\em JHEP}              {\bf #2}, {\em  #1},  #3}
\def\NPB#1#2#3{{\em Nucl.\ Phys.}\/ {\bf #2}, {\em B#1},  #3}
\def\PLB#1#2#3{{\em Phys.\ Lett.}\/ {\bf #2}, {\em B#1},  #3}
\def\PLA#1#2#3{{\em Phys.\ Lett.}\/ {\bf #2}, {\em A#1},  #3}
\def\PRD#1#2#3{{\em Phys.\ Rev.}\/  {\bf #2}, {\em D#1},  #3}
\def\PRL#1#2#3{{\em Phys.\ Rev.\ Lett.}\/ {\bf #2}, {\em #1}, #3}
\def\PRT#1#2#3{{\em Phys.\ Rep.}\/  {\bf #2}, {\em  #1},  #3}
\def\MPL#1#2#3{{\em Mod.\ Phys.\ Lett.}\/ {\bf #2}, {\em A#1}, #3}

\def\IJMP#1#2#3{{\em Int.\ J.\ Mod.\ Phys.}\/ {\bf #2}, {\em A#1}, #3}

\def\EJP#1#2#3{{\em Eur.\ Phys.\ Jour.}\/ {\bf #2}, {\em C#1}, #3}
\def\etal{{\it et. al.\/}~}

\lastpage{x}
\doinum{10.3390/------}
\pubvolume{2}
\pubyear{2014}
\history{Received: 28 February 2014; in revised form: 10 April 2014 / Accepted: 15 April 2014 / \\Published: 2014}
\pdfoutput=1

\usepackage{graphicx}
\usepackage{epsf}

\Title{String Phenomenology: Past, Present and Future Perspectives}

\Author{Alon E. Faraggi$^{1,}$*}

\address{%
$^{1}$ Department of Mathematical Sciences, University of Liverpool,
Liverpool L69 7ZL, United Kingdom}

\corres{alon.faraggi@liv.ac.uk.}

\abstract{
The observation of a scalar resonance at the LHC, compatible with 
perturbative electroweak symmetry breaking, reinforces the 
Standard Model parameterisation of all subatomic data. 
The logarithmic evolution of the SM gauge and matter 
parameters suggests that this parameterisation remains viable
up to the Planck scale, where gravitational effects are 
of comparable strength. String theory 
provides a perturbatively consistent scheme to explore how 
the parameters of the Standard Model may be determined from a
theory of quantum gravity. 
The free fermionic heterotic string models
provide concrete examples
of exact string solutions that reproduce the spectrum 
of the Minimal Supersymmetric Standard Model. 
Contemporary studies entail 
the development of methods to classify large classes of 
models. This led to the
discovery of exophobic heterotic--string vacua and the
observation of spinor--vector duality, which provides an insight 
to the global structure of the space of (2,0) heterotic--string vacua. 
Future directions entail the study of the role of the massive 
string states in these models and their incorporation in cosmological 
scenarios. A complementary direction is the formulation of 
quantum gravity from the principle of manifest phase space duality and 
the equivalence postulate of quantum mechanics, which suggest that
space is compact. The compactness of space, which implies intrinsic
regularisation, may be tightly related to the intrinsic finite length scale,
implied by string phenomenology. 
}


\begin{document}


\section{Introduction}

The experimental observation of a scalar resonance by the ATLAS \cite{atlas}
and CMS \cite{cms} experiments of the
Large Hadron Collider at CERN, compatible with the 
scalar particle of the Standard Electroweak Model \cite{weinberg},
is a pivotal moment in the quest for the 
unification of the fundamental theories of matter and interactions. 
Indeed, nearly thirty years have elapsed since the experimental discovery
of the $W^\pm$ and $Z$--vector bosons \cite{wpmzdiscovery,wpmzdiscovery2},
and forty years since the demonstration of renormalizability
of spontaneously broken non--Abelian gauge symmetries \cite{thooft}, 
which were the earlier milestones on this journey.
The discovery of the Higgs boson solidifies the Standard Model
parameterisation of all subatomic experimental observations
to date. The observation of a Higgs boson at 125GeV suggests that
the electroweak symmetry breaking mechanism is perturbative, rather than
nonperturbative. This reinforces the view that the Standard Model
provides a viable perturbative parameterisation of the subatomic 
interactions up to an energy scale, which is separated by orders
of magnitude from the scale within reach of contemporary accelerator
experiments. If this is indeed the scenario selected by nature, it entails 
that alternative experimental tests will be required to establish its validity.
These tests will unavoidably look for astrophysical and cosmological
imprints, that can probe the much higher energy scales.

The possibility that the Standard Model provides a viable effective
parameterisation, up to a much higher scale, has been 
entertained in the context of Grand Unified Theories (GUTs) and string 
theories \cite{ibanezuranga}.
The gauge charges of the Standard Model matter states 
are strongly suggestive of the embedding 
of the Standard Model states in representations of larger gauge 
groups. This is most striking in the context of $SO(10)$ GUT, 
in which each of the Standard Model chiral generations fits into a
${\bf 16}$ spinorial representation of $SO(10)$. The gauge charges of the 
Standard Model matter states are experimental observables. The Standard Model 
contains three generations, which are split into six multiplets that
are charged under its three gauge sectors. Therefore, in the framework
of the Standard Model one needs fifty--four parameters to account for these 
gauge charges. Embedding the Standard Model in $SO(10)$ reduces this
number of parameters to one parameter, which is the number of spinorial
${\bf 16}$ representations of $SO(10)$ needed to accommodate the Standard Model
spectrum. Additional evidence for the high scale unification stems from: 
\begin{itemize}
\item  the logarithmic running of the Standard Model parameters, 
which is compatible with observations in the gauge sectors
\cite{cwogcu} and 
the heavy generation Yukawa couplings \cite{cwohms}.
Logarithmic running 
in the scalar sector is spoiled by radiative corrections from the 
Standard Model cutoff scale. Restoration of the logarithmic 
running mandates the existence of a new symmetry. 
Supersymmetry is a concrete example that fulfils the task. 
The observation of a scalar resonance 
at 125GeV, and the fact that no other particles have been observed
up to the multi--TeV energy scale, indicates that the 
resonance is a fundamental scalar rather than a composite state
\cite{nussinov}. This outcome agrees with the Higgs states in
heterotic--string vacua. 
\item
Further evidence for the validity 
of the renormalizable Standard Model up to a very high 
energy scale stems from the suppression of proton decay mediating 
operators. The Standard Model 
should be regarded as providing a viable effective parametrisation,
but not as a fundamental accounting of the observable phenomena.
The reason being in that it does not provide a complete description. 
Obviously, gravitational effects are not accounted for.
Moreover, the Standard Model itself is not mathematically self--consistent. 
It gives rise to singularities in the ultraviolet limit. For these 
reasons the Standard Model can only be regarded as an effective theory 
below some cutoff. A plausible cutoff is the Planck scale, at which 
the gravitational coupling is of comparable strength to the gauge 
couplings. The renormalizability of the Standard Model is not 
valid beyond its cutoff scale. Nonrenormalizable operators are 
induced by whatever theory extends the Standard Model at and beyond 
the cutoff scale. We should therefore take into account
all the nonrenormalizable operators that are allowed 
by the Standard Model gauge symmetries, and that are suppressed
by powers of the cutoff scale. Such dimension six operators, 
which are invariant under the Standard Model gauge symmetries,
lead to proton decay. They indicate that the cutoff scale 
must be above $10^{16}$GeV, unless they are forbidden by some 
new symmetries. As global symmetries are, in general, expected
to be violated by quantum gravity effects, the new symmetries 
should be either gauge symmetries or local discrete symmetries \cite{ps}.
\item
Suppression of left--handed neutrino masses is compatible
with the generation of heavy mass to the right--handed
neutrinos by the seesaw mechanism \cite{seesaw}. 
\end{itemize}

The Standard Model multiplet structure, and the additional 
evidence provided by logarithmic running, proton longevity and
neutrino masses indicates that the primary guides in the
search of a realistic string vacuum are the existence of
three chiral generations and their embedding in $SO(10)$
representations.
This embedding does not entail the existence of an $SO(10)$ gauge 
symmetry in the effective low energy field theory. 
Rather, the $SO(10)$ symmetry is broken at the string
level to a maximal subgroup, and preferably directly 
to the Standard Model gauge group.

The Standard Model of particle physics is founded on a causal and 
renormalizable quantum field theory with local phase 
invariance under a product of Abelian and non--Abelian gauge symmetries.
These symmetry principles encode all the subatomic experimental 
observations to date. Alas, the effects of the gravitational 
interactions are not included in this picture. Moreover, 
there is a fundamental dichotomy between the principles underlying quantum
mechanics and gravitational observations. In particular,
with regard to the treatment of the vacuum. While quantum field theories
give rise to energy sources that contribute to the vacuum energy 
with scale of the order of the QCD scale and above, observations
show that the vacuum energy is smaller by orders of magnitude.
Another point of contention is with regard to the nature of space. 
In general relativity, the contemporary theory of gravity, 
space is a dynamical field satisfying Einstein's equations of motion. 
In quantum field theories, on the other hand, space provides background
parameters and does not correspond to the fundamental degrees of freedom, 
which are encoded in the particle wave--functions and their conjugate 
momenta. Furthermore, gravity as a quantum field theory is not renormalizable,
which is therefore plagued with infinities and is inconsistent at a 
fundamental level. 

The conundrum may be seen to arise from the fact that quantum field theories
may, in principle, probe space distances that are infinitely small, provided 
that the corresponding momenta is infinitely large. We may envision that this
outcome is fundamentally unphysical and what we need is a fundamental 
description of matter and interactions, which excludes the possibility of
probing infinitely small distances. String theory provides such a theory.
Moreover, the equivalence postulate formulation of quantum
mechanics implies that space is compact and the existence of a fundamental
length in quantum mechanics \cite{tqc}. The fundamental cutoff may therefore be
intrinsically
built into quantum mechanics, provided that its full set of symmetries are 
incorporated. 

As a finite theory string theory provides a 
consistent framework for perturbative quantum gravity
\cite{stringreviews}.
The consistency of string theory at the quantum level 
dictates that it must accommodate a specific number of 
worldsheet degrees of freedom to produce an anomaly free 
and finite theory. Some of degrees of freedom give rise to the
gauge symmetries that we may identify with the subatomic 
interactions. Moreover, similar consistency constraints at the
quantum level in the case of the superstring and heterotic--string
give rise to matter states that are charged under the gauge 
degrees of freedom, and may be identified with the Standard Model
matter states. Hence, string theory provides a viable framework
for the consistent unification of gravity with the subatomic 
matter and interactions. In turn, this feature of string
theory allows for the development of a phenomenological 
approach to quantum gravity.

String theory is therefore a mundane extension of the idealisation 
of point particles with internal attributes. Furthermore, 
the rank of the gauge group accounting for the internal attributes
is fixed by the consistency conditions of the theory. The string
action is parameterised by two worldsheet degrees of freedom, 
corresponding to the proper time and the string internal dimension. 
The equation of motion of the worldsheet degrees of freedom 
is a two dimensional wave equation. The solutions are separated 
into left-- and right--moving solutions. The 
physical states of the quantised string give 
rise to a tachyonic state, which is eliminated from the spectrum
if the bosonic worldsheet fields are augmented with fermionic 
fields. This is achieved provided that the theory possesses $N=2$ 
supersymmetry on the worldsheet, which guarantees the existence
of $N=1$ spacetime supersymmetry. Since the tachyonic state does not 
have a corresponding fermionic superpartner, the existence of spacetime
supersymmetry guarantees that the tachyonic state is excluded from 
the physical spectrum. Additionally, the fermionic string
gives rise to spacetime fermions that transforms in 
representations of the internal gauge symmetry. 

String theory is formulated as a perturbative scattering expansion. 
Using the conformal symmetry on the worldsheet, 
the lowest order amplitudes can be mapped to the sphere with 
vertex operator insertions corresponding to the external string states. 
Higher order amplitudes are mapped to higher genus tori, with the genus
one torus being the lowest order quantum correction.
The vacuum to vacuum amplitude is the first order correction 
when there are no external states and all the physical states
can propagate in the closed time--like loop. 
The conformal worldsheet symmetry is translated to 
invariance of the torus amplitude under modular transformations
of the complex worldsheet parameter $\tau$.  
The worldsheet fermionic fields can pick up non trivial 
phases when parallel transported around the non--contractible loops of 
the worldsheet torus. The possible transformations for all the 
worldsheet fermions are encoded in the so called spin structures
and are mixed non--trivially by the modular transformations. 
Requiring invariance under modular transformations leads to a set of
non--trivial constraints on the allowed spin structures
\cite{stringreviews}.

Different string theories may be formulated depending on the 
existence, or not, of worldsheet fermionic fields in the 
left-- and right--moving sectors of the string. Type IIA and
type IIB superstring arise if worldsheet fermions are added in
both the left-- and right--moving sectors. Adding worldsheet fermions
only to the left--moving sector produces the heterotic--string
with $E_8\times E_8$, or
$SO(32)$ gauge symmetry in ten dimensions. 
In the low energy point particle approximation, 
we expect a string theory to correspond to an effective
field theory approximation. That is when the energy involved
is not sufficiently high to reveal the internal structure of the 
string, we expect that it should be described effectively as
some point particle field theory. In the case of the fermionic
strings these are type IIA or IIB supergravities, or an effective
ten dimensional supergravity with $E_8\times E_8$ or $SO(32)$ 
gauge symmetry. Additionally, the non--perturbative effective field 
theory limits of the ten dimensional string are related to
compactifications of eleven dimensional supergravity. For example, 
the type IIA superstring is related to compactification of 
eleven dimensional supergravity on a circle, whereas the 
ten dimensional heterotic $E_8\times E_8$ string
corresponds to compactification
on a circle moded by a $Z_2$ reflection symmetry. The full set of
relations at the quantum level is yet to be unravelled, 
and is traditionally dubbed as M--theory or F--theory 
\cite{stringreviews}. 

The lesson is that our understanding of the 
synthesis of gravity and the gauge interactions is still
very rudimentary. String theory is clearly a step in the 
right direction. It provides a framework
to ask questions about the gauge and gravity unification and
to seek consistent answers within that framework. By giving
rise to all the basic fields that are used to parameterise 
the subatomic and gravitational experimental data, it enables 
the development of a phenomenological approach to quantum gravity. 
However, its is clear that string theory is not the final answer. 
The contemporary string theories are believed to be 
effective limits of a more fundamental theory. From that perspective
each of the string theories can be used to probe some properties 
of the vacuum of the fundamental theory, but not to fully characterise it. 
The heterotic $E_8\times E_8$ string is the effective 
limit that gives rise to spinorial $SO(10)$ representation 
in the perturbative spectrum. The heterotic--string therefore 
is the effective limit that should be used if the 
properties that we would like to preserve are the 
existence of three chiral generations and their 
embedding in spinorial $SO(10)$ representations. 

\section{Past}\label{past}

Realistic string models are obtained by compactifying the 
heterotic--string \cite{heterotic} from ten to four dimensions \cite{chsw}.
Alternatively, we can construct realistic string models
directly in four dimensions by representing the 
compactified dimensions in terms of internal conformal field theories
propagating on the string worldsheet. 
The simplest such theories are in terms of free worldsheet 
field theories, {\it i.e.} in terms of free bosons \cite{orbifolds}
or free fermions \cite{fff}, with the main simplification being the 
implementation of the modular invariance constraints. 
Nevertheless, constructions using interacting worldsheet 
conformal field theories exist as well \cite{gepner}, and can be used 
to construct phenomenological vacua \cite{grs}. It should be remarked that 
the representations of the four dimensional string vacua as
compactifications on internal manifold or in terms of internal 
conformal field theories are not necessarily distinct. For 
example, theories that utilise 
two dimensional worldsheet free bosons or free fermions 
are mathematically equivalent. Similarly, it was demonstrated
in some cases that string models with interacting internal CFT
correspond to string compactification on a Calabi--Yau manifold 
at specific points in the moduli space \cite{gepner}. This is an important
point for the following reason. While the space of distinct 
string vacua in the effective field theory limit may seem to be
huge, many of these vacua are related by various perturbative and
nonperturbative dualities at the string level. The reason is that
at the string level massless and massive
physical states can be exchanged. Thus, vacua that are topologically
and physically distinct in the effective field theory level are in
fact connected at the string level. This feature is particularly
important if we envision the existence of a dynamical vacuum 
selection mechanism in string theory. 

The simplest phenomenological string models can therefore be constructed
by using a free internal conformal field theory. String theories
in which the internal CFT is written in terms of free fermions 
corresponds to compactifications on a flat six dimensional torus
at a special point in the moduli space \cite{n4n2n1}.
Exactly marginal deformations
from the free fermionic point are obtained by adding Thirring
interactions among the worldsheet fermions \cite{chang}. 
The number of allowed deformations correspond exactly to the number
of allowed deformations in compactifications of the 
corresponding string theory on a flat torus. Compactifications
of the heterotic--string on a flat six dimensional torus produce 
$N=4$ spacetime supersymmetry, which is reduced to $N=1$ 
by modding out the internal six dimensional torus by an internal 
symmetry. This produces the so called orbifold compactifications. 
The simplest such orbifolds correspond to modding out the 
internal six dimensional torus by $Z_2$ symmetries. Modding out
by a single $Z_2$ reduces the number of spacetime supersymmetries
from $N=4$ to $N=2$. Therefore, to reduce the number of supersymmetries
to $N=1$ necessitates modding out by two independent $Z_2$ symmetries, 
{\it i.e.} by $Z_2\times Z_2$.

\subsection{NAHE--based models}

In the free fermionic formulation \cite{fff}
of toroidal compactifications \cite{narain,nsw}
all the internal degrees of freedom needed to cancel the worldsheet 
conformal anomaly are represented in terms of free fermions propagating 
on the string worldsheet. In the usual notation the 64 worldsheet 
fermions in the light--cone gauge are denoted as:
\leftline{~~~${\underline{{\hbox{Left-Movers}}}}$:~~~~~~~~~~~~~~~~~~~~~~~~
~~~~$\psi^\mu,~~{ \chi_i},~~{ y_i,~~\omega_i}~~~~(\mu=1,2,~i=1,\cdots,6)$}
\vspace{4mm}
\leftline{~~~${\underline{{\hbox{Right-Movers}}}}$}
$${\bar\phi}_{A=1,\cdots,44}=
\begin{cases}
~~{ {\bar y}_i~,~ {\bar\omega}_i} & i=1,{\cdots},6\cr
  & \cr
~~{ {\bar\eta}_i} & i=1,2,3~~\cr
~~{ {\bar\psi}_{1,\cdots,5}} & \cr
~~{{\bar\phi}_{1,\cdots,8}}  & 
\end{cases}
$$
In this notation the $\psi^{1,2}, \chi^{1,\cdots,6}$ are the fermionic 
superpartners of the left--moving bosonic coordinates. The 
$\{y,\omega\vert{\bar y},{\bar\omega}\}^{1,\cdots,6}$
are the worldsheet real fermions corresponding to the 
six compactified dimensions of the internal manifold.
The remaining sixteen complex fermions generate the Cartan
subalgebra of the ten dimensional gauge group, 
with ${\bar\psi}^{1,\cdots,5}$ being those that generate the $SO(10)$ 
symmetry, and ${\bar\phi}^{1,\cdots,8}$ are those that generate the 
hidden sector gauge group. The ${\bar\eta}^{1,2,3}$ complex worldsheet 
fermions generate three $U(1)$ symmetries.

Under parallel transport around the noncontractible loops of the 
torus amplitude the worldsheet fermionic fields can pick 
up a phase. The 64 phases are encoded in boundary condition basis vectors, 
which generate the one loop partition function, 
$${Z~=~\sum_{all~ spin\atop structures}~%
c{{\vec\xi}\choose{\vec\beta}}~
Z{{\vec\xi}\choose{\vec\beta}}},$$
where ${\vec\xi}$ and ${\vec\beta}$ denote all possible combinations 
of the basis vectors. 
The requirement of modular invariance leads to a set of constraints on the
allowed basis vectors and one loop phases. 
The basis vectors generate a finite additive group $\Xi$ and each sector
in the additive group, $\xi\in\Xi$, produces a Fock space by
acting on the vacuum with fermionic and bosonic oscillators.
Worldsheet fermionic fields that are periodic under parallel transport
produce a doubly degenerate vacuum that generate the spinorial charges. 
The physical states in the Hilbert space are obtained by applying the 
Generalised GSO projections, which arise due to the
modular invariance requirement. 
The cubic level and higher order terms in the superpotential are 
obtained by calculating scattering amplitudes between vertex operators
\cite{superpoterm}.
Finally, string vacua often give rise to an pseudo--anomalous $U(1)$ 
symmetry, which is cancelled by the Green--Schwarz mechanism 
\cite{gs,dsw}. 
The anomalous $U(1)$ gives rise to a Fayet--Iliopoulos $D$--term
\cite{dsw,ads}, which breaks spacetime supersymmetry at the string scale.
Restoration of supersymmetry is obtained by assigning non--trivial
Vacuum Expectation Value (VEV) to a set of fields in the physical
spectrum, and imposing that all the supersymmetry breaking
$F$-- and $D$--terms vanish. 

In this manner a large set of string vacua can be obtained. 
The early quasi--realistic free fermionic models were constructed
in the late 1980s -- early 1990s, and consist of the so called 
NAHE--based models. The NAHE--set is a set of five boundary 
condition basis vectors, $\{{\bf1},S,b_1,b_2,b_3\}$, 
which is common to a large class of the early models \cite{nahe}. 
The two basis vectors $\{{\bf1},S\}$ correspond to a toroidally 
compactified model with $N=4$ spacetime supersymmetry and an 
$SO(44)$ gauge group. The sectors $b_1$, $b_2$ and $b_3$
correspond to the three twisted sectors of a $Z_2\times Z_2$ 
orbifold compactification. They reduce the number of 
supersymmetries to $N=1$ and the gauge symmetry to 
$SO(10)\times SO(6)^3\times E_8$. Additionally, they produce 
48 multiplets in the spinorial ${\bf 16}$ representation of $SO(10)$.
The number of these multiplets is reduced to three by adding 
three additional basis vectors to the NAHE set, typically denoted 
by $\{\alpha, \beta, \gamma\}$, which also reduce the gauge symmetry. 
The $SO(10)\times E_8$ symmetry is reduced to a maximal subgroup and 
the flavour $SO(6)^3$ symmetries are reduced to $U(1)^n$, 
with $n=3,\cdots, 9$. Using this construction three generation 
models with 
\begin{itemize}
\item $SU(5)\times U(1)$ \cite{fsufive}; 
\item $SU(3)\times SU(2)\times U(1)^2$ \cite{fny,eu,slm}; 
\item $SO(6)\times SO(4)$ \cite{alr} and
\item $SU(3)\times SU(2)^2 \times U(1)$ \cite{lrs}, 
\end{itemize}
were obtained, whereas models with $SU(4)\times SU(2) \times U(1)$ \cite{su421}
did not yield three generations. It is noted that in all
these models the Standard Model weak hypercharge possess the 
$SO(10)$ embedding, and yield the canonical GUT normalisation
$\sin^2\theta_W(M_S)=3/8$, where $M_S$ is the string unification
scale. This is an important feature of these
models because it facilitates agreement with the measured 
gauge coupling parameters at the electroweak scale \cite{df}.
It should also be contrasted with other possible embedding 
of the weak hypercharge that do not yield the canonical 
GUT embedding. Such is the case, for instance, in many 
orientifold models. However, in orientifold models the string 
scale may be lowered relative to the gravitational scale. 
Hence, in orientifold models smaller values of $\sin^2\theta_W(M_S)$
may be accommodated. Heterotic--string models may also 
yield smaller values of $\sin^2\theta_W(M_S)$, by modifying the 
identification of the weak hypercharge in the string models. 
We recall that $\sin^2\theta_W(M_S)$ arises as a 
result of the relative normalisation of the weak hypercharge 
relative to the non--Abelian generators at $M_S$ \cite{dfm}.
In heterotic string 
models this normalisation is affected by the number of Cartan 
generators in the weak hypercharge combination relative to the 
number of Cartan subgenerators of the non--Abelian group factors. 
However, in the perturbative heterotic string the unification scale is fixed
and therefore lower values of $\sin^2\theta_W(M_S)$ are disfavoured. 
This constraint may be relaxed in the non--perturbative heterotic--string
\cite{witten}.
Another point to note in regard to the definition of the 
weak hypercharge is the existence of string states that carry fractional 
electric charge. This is a general feature of string models. The reason
being the breaking of the non--Abelian gauge symmetries by Wilson lines. 
A general observation by Wen and Witten \cite{wenwitten}, and a theorem 
by Schellekens \cite{schellekens}, notes that breaking a non--Abelian gauge
group by a Wilson line in string theory, with a left over 
unbroken $U(1)$ symmetry, produces states that do not satisfy the $U(1)$ charge
quantisation of the unbroken non--Abelian symmetry. This outcome
further depends on the identification of the weak--hypercharge. 
That is if we relax the canonical GUT embedding of the weak 
hypercharge, we modify the GUT quantisation of the 
$U(1)$ charges and may therefore obtain integrally charged states. 
The important point to note is that these are phenomenological properties
of string constructions and it is yet to be determined how they 
play out in fully realistic string constructions. 

\subsection{Phenomenology of string unification}

Subsequent to the construction of the string models and analysis 
of their spectra, we calculate the cubic level and higher order
terms in the superpotential, up to a desired order 
for a specific phenomenological problem. The next step entails
the analysis of supersymmetric $F$ and $D$--flat directions.
Requiring that the vacuum at the string scale is supersymmetric
necessitates the assignment of non--vanishing VEVs to a set of Standard
Model singlets in the string models. In this process some of 
the higher order terms in the superpotential become effective renormalizable 
operators, which are suppressed relative to the leading order cubic terms, 
{\it i.e.} 
\begin{equation}
{V_1^fV_2^fV_3^b\cdot\cdot\cdot\cdot V_N^b}~\rightarrow~%
V_1^fV_2^fV_3^b{{\langle V_4^b\cdots V_N^b\rangle}\over{M^{N-3}}},
\label{superpot} 
\end{equation}
where $V^{f,b}$ are fermionic and bosonic vertex operators, respectively;
$N$ is the order of the nonrenormalisable operator; and $M$ is the 
string cutoff scale.  
Using this methodology many of the issues pertaining to the phenomenology
of the Standard Model and string unification have been studied in the 
framework of the quasi--realistic free fermionic heterotic--string
models. A partial list includes: 
\begin{itemize}
\item [{$\bullet$}] 
$\underline{\hbox{Top quark mass prediction.}}$
The analysis of fermion 
masses entails the calculation of cubic and higher order terms 
in the superpotential that are reduced to dimension four terms 
in eq. (\ref{superpot}). The Standard Model fermion mass terms 
arise from couplings to the electroweak Higgs with an assumed VEV
of the order of the electroweak scale. Other fermion mass terms 
arise from coupling to other scalar fields, and their mass scales
may therefore be higher than the electroweak scale. Analysis 
of Standard Model fermion masses yielded a viable prediction for
top quark mass prior to its experimental observation \cite{top}.
The calculation proceeds as follows. First the top quark Yukawa 
coupling is calculated at the cubic level of the superpotential,
giving $\lambda_t=\langle Q t_L^c H\rangle = \sqrt2 g$, 
where $g$ is the gauge coupling at the unification scale.
Subsequently, the Yukawa couplings for the bottom quark and tau lepton 
are obtained from quartic order terms. The magnitude of the quartic order
coefficients are calculated using standard CFT techniques, and the 
VEV of the Standard Model singlet field in the relevant terms is 
extracted from analysis of the $F$-- and $D$--flat directions.
This analysis yields effective Yukawa couplings for the 
bottom quark and tau lepton in terms of the unified gauge coupling 
given by $\lambda_b=\lambda_\tau=0.35 g^3\sim~{1/8}\lambda_t$ \cite{top}.
This result for the top quark Yukawa coupling 
is common in a large class of free fermionic models, 
whereas those for the bottom quark and tau lepton 
differ between models. Similarly, the Yukawa coupling
for the two lighter generations differ
between models and depend on the flat direction VEVs.
Subsequent to extracting the Yukawa couplings at the string 
scale, they are run to the electroweak scale using the
Minimal Supersymmetric Standard Model (MSSM) Renormalisation
Group Equations (RGEs). It is further assumed that the unified 
gauge coupling at the string scale is compatible with the value 
required by the gauge coupling data at the electroweak scale. 
The bottom Yukawa is further run to the
bottom mass scale, which is used to extract a value for $\tan\beta=v_1/v_2$,
with $v_1$ and $v_2$ being the VEVs of the two MSSM electroweak Higgs 
doublets. The top quark mass is then given by
$$m_t~=~\lambda_t(m_t)~{{{v}_0}\over{\sqrt2}}~{\tan\beta\over
{(1+\tan^2\beta)^{1\over2}}}$$
with $v_0=\sqrt{2(v_1^2+v_2^2)}=246{\rm GeV}$, yielding
$m_t\sim175-180{\rm GeV}$. It is noted that, up to the assumptions
stated above, the top Yukawa coupling is found near a fixed point. 
Namely, varying the top Yukawa between $0.5-1.5$ at the unification
scale yields $\lambda_t(M_Z)\sim1$ at the electroweak scale.
This calculation demonstrates the important
advantage of string theory over other attempts of
developing a viable framework for quantum gravity. It unifies
the gauge and Yukawa couplings 
and enables the calculation of the Standard Model
Yukawa couplings in terms of the unified string coupling. While 
the calculation of the top Yukawa is robust and shared between
a large class of models,  the calculation of the corresponding
couplings for the lighter quarks and leptons involve a large degree
of model dependence. Before investing
substantial efforts to calculate the Yukawa couplings of the lighter 
quarks and leptons in a given model, we should enhance the prospect
that a given model is the right model. This line of reasoning 
underlies the contemporary approach that is outlined below. 

\item[{$\bullet$}] 
$\underline{\hbox{Standard Model fermion masses.}}$
The analysis of the effective Yukawa couplings for the lighter
two generations proceeds by analysing higher order terms 
in the superpotential and extracting the effective
dimension four operators \cite{superpoterm}. The analysis should be regarded
as demonstrating in principle the potential of string models
to explain the detailed features of the Standard Model
flavour parameters. It is still marred by too many uncertainties 
and built in assumptions to be regarded as a predictive framework. 
Nevertheless, once an appealing model is constructed 
the methodology is in place to attempt a more predictive
analysis. The explorations to date included, for example, 
demonstration of the generation mass hierarchy \cite{gmh};
Cabibbo--Kobayashi--Maskawa (CKM) mixing \cite{fh1,fh3};
light generation masses \cite{lgm}; and neutrino masses 
\cite{fh2,cfneutrinomasses}.

\item[{$\bullet$}]
$\underline{\hbox{Gauge coupling unification}.}$ 
An important issue in heterotic--string models is compatibility 
with the experimental gauge coupling data at the electroweak scale. 
The perturbative heterotic--string predicts that the gauge
couplings unify at the string scale, which is of the order of 
$5\times 10^{17}{\rm GeV}$. On the other hand extrapolation 
of the gauge couplings, assuming MSSM spectra, from the 
$Z$--boson mass scale to the GUT scale shows that the 
couplings converge at a scale of the order of 
$2\times 10^{16}{\rm GeV}$. Thus, the two scales differ
by a factor of about 20.
This extrapolation should be taken with caution as the 
the parameters are extrapolated over 14 orders 
of magnitude, with rather strong assumptions on the 
physics in the region of extrapolation. 
Indeed, in view of
the more recent results from the LHC the analysis needs to be revised
as the assumption of MSSM spectrum at the $Z$--boson scale
has been invalidated. Nevertheless, the issue can be 
studied in detail in perturbative heterotic--string models and 
a variety of possible effects have been examined, including
heavy string threshold corrections, light SUSY thresholds, 
additional gauge structures and additional intermediate matter
states \cite{df}. Within the context of the free fermionic models
only the existence of additional matter states may resolve the 
discrepancy and such states indeed exist in the spectrum of 
concrete string models \cite{gcu}. This result may be relaxed in the 
nonperturbative heterotic--string \cite{witten}
or if the moduli are away from the free fermionic point \cite{nillesstein}. 

\item[{$\bullet$}] 
$\underline{\hbox{Proton stability.}}$
Proton longevity is an important problem in quantum gravity, in general,
and in string models in particular. The reason being that we expect 
only gauge symmetries, or local discrete symmetries that arise as 
remnants of broken gauge symmetries, to be respected in quantum gravity. 
Within the Standard Model itself baryon and lepton are accidental global 
symmetries at the renormalizable level. 
Thus, we expect, in general, all operators that are compatible with
the local gauge and discrete symmetries in given string models to be
generated from nonrenormalizable terms. Such terms can then give rise to 
dimension four, five and six baryon and lepton number violating
operators that may lead to rapid proton decay. 
Possible resolutions have been studied in specific free fermionic 
models and include the existence of an additional light $U(1)$ symmetry
\cite{extrazprime} and local discrete symmetries \cite{ps}. 
 
\item[{$\bullet$}] 
$\underline{\hbox{Squark degeneracy.}}$
String models may, in general, lead to non-degenerate squark masses, 
depending on the specific SUSY breaking mechanism. For example, 
SUSY breaking mechanism which is dominated by the moduli $F$--term 
will lead to non--degenerate squark masses, because of the 
moduli dependence of the flavour parameters. Similarly, 
$D$--term 
SUSY breaking depends on the charges 
of the Standard Model fields under the gauge symmetry
in the SUSY breaking sector, and those are in general family 
non--universal. Free fermionic models can give rise to 
a family universal anomalous $U(1)$ \cite{auone}. If the SUSY breaking
mechanism is dominated by the anomalous $U(1)$ $D$--term 
it may produce family universal squark masses of order $1{\rm TeV}$
\cite{squarkdg}.

\item[{$\bullet$}]
$\underline{\hbox{Minimal Standard Heterotic--String Model (MSHSM).}}$
Three generation semi--realistic string models produce, in general, 
additional massless vector--like states that are charged under the 
Standard Model gauge symmetries. Some of these additional
vector--like states arise from the Wilson line breaking
of the $SO(10)$ GUT symmetry and therefore carry fractional
charge with respect to the remnant unbroken $U(1)$ symmetries. 
In particular, they may carry fractional electric charge, 
which is highly constrained by observations. These 
fractionally charged states must therefore be sufficiently massive
or diluted to evade the experimental limits. 
Mass terms for the vector--like states may arise from 
cubic and higher level terms in the superpotential.
In the model of ref. \cite{fny} its has been demonstrated in 
\cite{fractional} that all the exotic fractionally charged
states couple to a set of $SO(10)$ singlets. In ref. \cite{cfn}
$F$-- and $D$--flat solutions that incorporate this set 
of fields have been found. Additionally, all the extra 
standard--like fields in the model, beyond the MSSM, 
receive mass terms by the same set of VEVs. These solutions
therefore give rise to the first known string solutions that 
produce in the low energy effective theory of the observable sector
solely the states of the MSSM, and are dubbed 
Minimal Standard Heterotic--String Model (MSHSM).
Three generation Pati--Salam free fermionic models 
in which fractionally charged exotic states arise only 
in the massive spectrum were found in ref. \cite{exophobic}. 
Flat directions that lead to MSHSM with one leading Yukawa 
coupling were found in an exemplary model in this class \cite{cfr}.

\item[{$\bullet$}] 
$\underline{\hbox{Moduli fixing.}}$
An important issue in string models is that of moduli stabilisation. 
The free fermionic models are formulated near the self--dual 
point in the moduli space. However, the geometrical moduli that
allow deformations from that point exist in the spectrum and can 
be incorporated in the form of Thirring worldsheet interactions
\cite{chang}. The correspondence of the free fermionic 
models with $Z_2\times Z_2$ orbifolds implies that the geometrical 
moduli correspond to three complex and three K\"ahler structure 
moduli. String theory as a theory of quantum geometry, rather than
classical geometry, allows for assignment of asymmetric boundary
conditions with respect to the worldsheet fermions that correspond
to the internal dimensions. These correspond to the asymmetric bosonic
identifications under $X_L+X_R\rightarrow X_R-X_L$. In the free fermionic
models, and consequently in $Z_2\times Z_2$ orbifolds, it is possible 
to assign asymmetric boundary conditions with respect to six circles 
of the six dimensional compactified torus. In such a model 
all the complex and K\"ahler moduli of the untwisted moduli are projected 
out \cite{modulifixing}. Additionally, the breaking of the $N=2$
worldsheet supersymmetry in the bosonic sector of the heterotic--string
results in projection of the would--be twisted moduli \cite{modulifixing}. 
Thus, all the fields that are naively identified as moduli in models
with $(2,2)$ worldsheet supersymmetry can be projected out in 
concrete models. However, the identification of the moduli in models 
with $(2,0)$ worldsheet supersymmetry is not well understood and
there may exist other fields in the spectrum of such models that may 
be identified as moduli fields. Furthermore, as long as supersymmetry
remains unbroken in the vacuum there exist moduli fields associated with 
the supersymmetric flat directions. However, it has been
proposed that there exit quasi--realistic 
free fermionic models which do not admit supersymmetric flat directions
\cite{minimalhiggs}. This is obtained when both symmetric and 
asymmetric twistings of the internal dimensions are implemented, resulting
in reduction of the number of moduli fields. In the relevant models
supersymmetry is broken due to the existence of a Fayet--Iliopoulos term, which
is generated by an anomalous $U(1)$ symmetry. It was argued in
\cite{minimalhiggs} that the relevant models do not admit exact flat
directions and therefore supersymmetry is broken at some level. 
In such models all the moduli are fixed. It should be noted
that this possibility arises only in very particular string
models, rather than in a generic string vacua \cite{baylorgang}.
\end{itemize}

\section{Present}\label{present}

Most of the studies discussed so far were done by studying 
concrete examples of NAHE--based models, {\it i.e.} models 
that contain the common set $\{{\bf1},S,b_1,b_2,b_3\}$ plus the
three (or four) additional basis vectors $\{\alpha,\beta,\gamma\}$ 
that extend the NAHE--set and differ between models, with the 
most studied  models being those of ref. \cite{fny} and \cite{eu}.
More recent studies involve the exploration of large number of 
models. This provides an insight into the general properties 
of the space of vacua, as well as the development of a ``fishing
algorithm'' to fish models with specific phenomenological properties. 
This method led to 
discovery of spinor--vector duality \cite{spinvecdual} 
and of exophobic vacua \cite{exophobic,asseltwo,cfr,bfgrs}.
More recently the method has been applied for the 
classification of flipped $SU(5)$ free fermionic models \cite{frs}, 
as well as the classification with respect to the top quark Yukawa 
coupling \cite{johnstopclass}.

\subsection{Classification of fermionic $Z_2\times Z_2$ orbifolds}

Over the past decade a systematic method is being developed
that allows the explorations of large number of string vacua 
and analysis of their spectra. In this method the set of basis 
vectors is fixed. The Pati--Salam class of models is 
generated by a set of thirteen basis vectors
$$
B=\{v_1,v_2,\dots,v_{13}\},
$$
where
\begin{eqnarray}
v_1={\bf1}&=&\{\psi^\mu,\
\chi^{1,\dots,6},y^{1,\dots,6}, \omega^{1,\dots,6}| \nonumber\\
& & ~~~\bar{y}^{1,\dots,6},\bar{\omega}^{1,\dots,6},
\bar{\eta}^{1,2,3},
\bar{\psi}^{1,\dots,5},\bar{\phi}^{1,\dots,8}\},\nonumber\\
v_2=S&=&\{\psi^\mu,\chi^{1,\dots,6}\},\nonumber\\
v_{2+i}=e_i&=&\{y^{i},\omega^{i}|\bar{y}^i,\bar{\omega}^i\}, \
i=1,\dots,6,\nonumber\\
v_{9}=b_1&=&\{\chi^{34},\chi^{56},y^{34},y^{56}|\bar{y}^{34},
\bar{y}^{56},\bar{\eta}^1,\bar{\psi}^{1,\dots,5}\},\label{basis}\\
v_{10}=b_2&=&\{\chi^{12},\chi^{56},y^{12},y^{56}|\bar{y}^{12},
\bar{y}^{56},\bar{\eta}^2,\bar{\psi}^{1,\dots,5}\},\nonumber\\
v_{11}=z_1&=&\{\bar{\phi}^{1,\dots,4}\},\nonumber\\
v_{12}=z_2&=&\{\bar{\phi}^{5,\dots,8}\},\nonumber\\
v_{13}=\alpha &=& \{\bar{\psi}^{4,5},\bar{\phi}^{1,2}\}.
\nonumber
\end{eqnarray}
In the notation employed in Eq. (\ref{basis}) the worldsheet
fields appearing in a given basis vector have periodic 
boundary conditions, whereas all other fields have
anti--periodic boundary conditions.
The first twelve vectors in this set are identical to
those used in \cite{fknr,fkr} for the classification of
fermionic $Z_2\times Z_2$ orbifolds with $SO(10)$ GUT symmetry.
The thirteenth basis vector, $\alpha$, breaks the $SO(10)$ symmetry
and generates the Pati--Salam class of models. 
The set $\{1,S\}$ generate an
$N=4$ supersymmetric model, with $SO(44)$ gauge symmetry.
The vectors $e_i,i=1,\dots,6$ give rise
to all possible symmetric shifts of the six internal fermionized coordinates
($\partial X^i=y^i\omega^i, {\bar\partial} X^i= \bar{y}^i\bar{\omega}^i$).
Their addition breaks the $SO(44)$ gauge group, but preserves
$N=4$ supersymmetry.
The vectors $b_1$ and $b_2$  define the $SO(10)$ gauge symmetry and
the $Z_2\times Z_2$ orbifold twists, which break
$N=4$ to $N=1$ supersymmetry.
The $z_1$ and $z_2$ basis vectors reduce the untwisted gauge group generators
from $SO(16)$ to $SO(8)_1\times SO(8)_2$.
Finally $v_{13}$ is the additional new vector that breaks the $SO(10)$ GUT
symmetry to
$SO(6)\times SO(4)$, and the $SO(8)_1$ hidden symmetry to
$SO(4)_1\times SO(4)_2$.

The second ingredient that is needed to
define the string vacuum are the GGSO projection coefficients that
appear in the one--loop partition function,
$c{{vi}\atopwithdelims() {v_j}}~=~{\rm exp}[i\pi(v_i|v_j)]$, 
spanning a $13\times 13$ matrix.
Only the elements with $i>j$ are
independent while the others are fixed by modular invariance.
A priori there are therefore 78 independent coefficients corresponding
to $2^{78}$ string vacua. Eleven coefficients
are fixed by requiring that the models possess $N=1$ supersymmetry.
Additionally, the phase $\cc{b_1}{b_2}$ only affects the overall chirality. 
Without loss of generality the associated GGSO projection
coefficients are fixed, leaving 66 independent coefficients.
Each of the 66 independent coefficients can take two discrete
values $\pm1$ and thus a simple counting gives $2^{66}$
(that is approximately $10^{19.9}$) models in the
class of superstring vacua under consideration.

The utility of the classification method is that it provides 
the means to span all the massless producing sectors in the models. 
For example, the twisted matter states arise from the sectors
\begin{eqnarray}
{B_{\ell_3^1\ell_4^1\ell_5^1\ell_6^1}^1}&=&{S+b_1+\ell_3^1 e_3+\ell_4^1 e_4 +
\ell_5^1 e_5 + \ell_6^1 e_6} \nonumber\\
{B_{\ell_1^2\ell_2^2\ell_5^2\ell_6^2}^2}&=&{S+b_2+\ell_1^2 e_1+\ell_2^2 e_2 +
\ell_5^2 e_5 + \ell_6^2 e_6} \nonumber\\
{B_{\ell_1^3\ell_2^3\ell_3^3\ell_4^3}^3}&=&
{S+b_3+ \ell_1^3 e_1+\ell_2^3 e_2 +\ell_3^3 e_3+ \ell_4^3 e_4}\nonumber
\end{eqnarray}
where ${l_i^j~=~0,1}$, 
$b_3=b_1+b_2+x=1+S+b_1+b_2+\sum_{i=1}^6 e_i+\sum_{n=1}^2z_n$ and
$x$ is given
by the vector $x=\{{\bar\psi}^{1,\cdots,5},{\bar\eta}^{1,2,3}\}$.
These sectors give rise to ${\bf16}$ and $\overline{\bf 16}$
representations of $SO(10)$ decomposed under
$SO(6)\times SO(4)\equiv SU(4)\times SU(2)_L\times SU(2)_R$.
The important feature of this classification method is that 
each of the sectors $B_{\ell_1^i\ell_2^i\ell_3^i\ell_4^i}$ for given 
${\ell_1^i\ell_2^i\ell_3^i\ell_4^i}$ gives rise 
to one spinorial, or one anti--spinorial, or neither, {\it i.e}
the states arising at each fixed point of the corresponding 
$Z_2\times Z_2$ are controlled individually. Similarly, 
the states from the sectors 
$
B_{\ell_1^i\ell_2^i\ell_3^i\ell_4^i}+x~
$
produce states in the vectorial ${\bf10}$ representation of $SO(10)$
decomposed under the Pati--Salam gauge group.

The power of the free fermionic classification method is that
it enables translation of the GGSO projections into generic 
algebraic forms. The general expression for the GSO 
projections on the states from a given sector $\xi\in\Xi$ is 
given by \cite{fff}
$$
{\rm e}^{i\pi (v_j\cdot F_\xi)}\vert S\rangle_\xi =
\delta_\xi\cc{\xi}{v_j}^*\vert S\rangle_\xi ~.
$$
From this expression we note that, whenever the overlap of periodic
fermions between the basis vector $v_j$ and the sector $\xi$ is
empty, the operator on the left of this expression is fixed. Hence, 
depending on the choice of the GGSO phase on the right, the 
given state is either in or out of the physical spectrum. 
For any given state from specific sectors there are several
basis vectors that act as projectors. Introducing the notation
$\cc{a_i}{a_j}=\exp{(a_i\vert a_j)}$ with $(a_i\vert a_j)=0,1$,
we can collect these projectors into algebraic system of equations
of the form 
$
\Delta^{(i)}\,U_{16}^{(i)}=Y_{16}^{(I)}\ ,\ i=1,2,3,
$
~where the unknowns are the fixed point labels
$
U_{16}^{(i)}=
\left[
p_{16}^i, 
q_{16}^i, 
r_{16}^i, 
s_{16}^i
\right]. 
$
~The $\Delta^i$ and $Y_{16}^i$ are given in terms of the GGSO projection
coefficients for each of the three planes. For example, on the first
plane for the spinorial ${\bf16}$ or $\overline{\bf 16}$ states we have
\begin{equation}
\Delta^{(1)}=\left[
\begin{array}{cccc}
\oo{e_1}{e_3}&\oo{e_1}{e_4}&\oo{e_1}{e_5}&\oo{e_1}{e_6}\\
\oo{e_2}{e_3}&\oo{e_2}{e_4}&\oo{e_2}{e_5}&\oo{e_2}{e_6}\\
\oo{z_1}{e_3}&\oo{z_1}{e_4}&\oo{z_1}{e_5}&\oo{z_1}{e_6}\\
\oo{z_2}{e_3}&\oo{z_2}{e_4}&\oo{z_2}{e_5}&\oo{z_2}{e_6}
\end{array}
\right]
\end{equation}
and
$
{Y_{16}^{(1)}=
\left[
\oo{e_1}{b_1},
\oo{e_2}{b_1},
\oo{z_1}{b_1},
\oo{z_2}{b_2}
\right]}
$
~with similar expressions for the second and third planes. 
The number of solutions per plane is determined by the 
relative rank of the matrix $\Delta^i$ and the rank
of the augmented matrix $(\Delta^i, Y_{16}^i)$. For
a given choice of GGSO projection coefficients, the number
of states surviving in the spectrum, is therefore readily 
obtained. 
Similar, algebraic expressions can be obtained 
for all the sectors that produce massless states in the 
given basis, as well as for the chirality of the fermions
with periodic boundary conditions.

The methodology outlined above enables the classification 
of a large number of fermionic $Z_2\times Z_2$ orbifolds. 
Compared to the earlier construction it enables a scan
of a large number of models and extraction of some of the
desired phenomenological properties. We can 
develop a fishing algorithm to extract models with 
specific characteristics. For example, a class of 
Pati--Salam models in which exotic fractionally charged
states appear as massive states but not in the massless
spectrum was found using these tools. The systematic classification
methods were developed to date only for models that admit 
symmetric boundary conditions with respect to the set of 
internal worldsheet fermions 
$\{y, \omega\vert{\bar y},{\bar\omega}\}^{1, \cdots, 6}$. 
On the other hand, NAHE--based models were constructed
using symmetric and asymmetric boundary conditions, with 
the assignment of asymmetric boundary conditions having
distinct phenomenological implications \cite{dtsm,udysr}.

\subsubsection{Spinor--vector duality}

Another example of the utility of the fermionic classification method
is given by the spinor--vector duality, which was discovered 
by using these methods and elucidates
the global structure of the free fermionic models, in 
particular, and that of the larger string landscape, 
in general. The spinor--vector duality is a duality in the
space of string vacua generated by the basis set $v_i$ with $i=1,\dots,12$, 
and unbroken $SO(10)$ symmetry. The duality
entails an invariance 
under the exchange of the total number of $({\bf16}+\overline{\bf16})$ representations
and the total number of ${\bf10}$ representations of $SO(10)$. That is,
for a given vacuum with a number of $({\bf16}+\overline{\bf16})$ 
and ${\bf10}$ representations, there exist another vacuum in which the
two numbers are interchanged. The origin of this duality is 
revealed when the $SO(10)$ symmetry is enhanced to $E_6$. 
Under the decomposition of $E_6\rightarrow SO(10)\times U(1)$ 
the ${\bf27}$ and $\overline{\bf27}$ representations decompose as 
${\bf27}={\bf16}+{\bf10}+{\bf1}$ and 
$\overline{\bf27}=\overline{\bf16}+{\bf10}+{\bf1}$. 
Therefore, in the case of vacua with $E_6$ symmetry 
the total number of $({\bf16}+\overline{\bf16})$ representations
is equal to the total of ${\bf10}$ representations. Hence,
models with enhanced $E_6$ symmetry are self--dual
under the spinor--vector duality map. 

The spinor--vector duality therefore arises from the 
breaking of the $E_6$ symmetry to $SO(10)\times U(1)$. 
This breaking is generated in the orbifold language by 
Wilson--lines, or in the free fermionic construction,
by choices of the GGSO projection coefficients. It is important
to recognise that these two descriptions are not distinct,
but are mathematically identical. That is we can translate
the GGSO projection coefficients to Wilson line and visa versa 
\cite{n4n2n1}.
Thus, when the $E_6$ symmetry is broken to $SO(10)\times U(1)$, 
there exist a choice of GGSO projection coefficients, 
or of Wilson lines, that keeps a number of spinorial 
$({\bf16}+\overline{\bf16})$ and a number of
vectorial ${\bf10}$ representations of $SO(10)$, 
and another choice for which the two numbers are interchanged.
It is important to note that this is an exact duality symmetry
operating in the entire space of string vacua in which the 
$SO(10)$ symmetry is not enhanced to $E_6$
\cite{spinvecdual,tristan,aft,ffmt}.
It is further noted that the spinor--vector duality can be 
interpreted in terms of a spectral flow operator \cite{ffmt}.
In this context the spectral flow operator in the twisted sector
may be seen as a deformed version of the operator inducing the 
Massive Spectral boson--fermion Degeneracy Symmetry (MSDS) \cite{msds}.
Therefore, the spinor--vector duality extends to the massive 
sectors \cite{ffmt}, 
albeit in a fashion that still needs to be determined in the general 
case. Similarly, we note that the generalisation 
of the spinor--vector duality to the case of interacting 
internal CFTs can be studied by adopting the following methodology
{\it e.g.} in the case of minimal models. The starting 
point is an heterotic--string compactified to four 
dimensions with $(2,2)$ worldsheet supersymmetry and 
an internal interacting CFT representing the compact space. 
The next step is to break the worldsheet supersymmetry
in the bosonic sector of the heterotic--string. 
The spectral flow operator then induces a map between
distinct $(2,0)$ vacua \cite{afg}.

We can also understand the spinor--vector duality operationally 
in terms of the free phases in the fermionic language \cite{tristan}
or as discrete torsion in the orbifold picture \cite{aft,ffmt}. 
For that purpose we recall the level one $SO(2n)$ characters
\cite{manno}
\beqn
O_{2n} &=& {1\over 2} \left( {\theta_3^n \over \eta^n} +
{\theta_4^n \over \eta^n}\right) \,,
~~~~~~~~~~~~
V_{2n} = {1\over 2} \left( {\theta_3^n \over \eta^n} -
{\theta_4^n \over \eta^n}\right) \,,
\nonumber \\
S_{2n} &=& {1\over 2} \left( {\theta_2^n \over \eta^n} +
i^{-n} {\theta_1^n \over \eta^n} \right) \,,
~~~~~~~
C_{2n} = {1\over 2} \left( {\theta_2^n \over \eta^n} -
i^{-n} {\theta_1^n \over \eta^n} \right) \,.
\nonumber
\eeqn
{where} 
\beq
{
\theta_3\equiv Z_f{0\choose0}~~~
  \theta_4\equiv Z_f{0\choose1}~~~}
{
  \theta_2\equiv Z_f{1\choose0}~~~
  \theta_1\equiv Z_f{1\choose1}~,~~}\nonumber
\eeq
and $Z_f$ is the partition function of a single
worldsheet complex fermion, given in terms of 
theta functions \cite{manno}. 
The partition function of the $E_8\times E_8$ heterotic--string
compactified on a six dimensional torus is given by
\beq
{Z}_+ = (V_8 - S_8) \, 
\left( \sum_{m,n} \Lambda_{m,n}\right)^{\otimes 6}\, 
\left(\bar O _{16} + \bar S_{16} \right) 
\left(\bar O _{16} + \bar S_{16} \right)\,,
\label{zplus}
\eeq
where as usual, for each circle,
\beq
p_{\rm L,R}^i = {m_i \over R_i} \pm {n_i R_i \over \alpha '} \,
~~~~~~~~~
{\rm and}
~~~~~~~~~
\Lambda_{m,n} = {q^{{\alpha ' \over 4} 
p_{\rm L}^2} \, \bar q ^{{\alpha ' \over 4} 
p_{\rm R}^2} \over |\eta|^2}\,.
\nonumber
\eeq
Next, a $Z_2\times Z_2^\prime:g\times g^\prime$ 
projection is applied, where 
the first $Z_2$ is a freely acting Scherk--Schwarz like projection, which 
couples a fermion number in the observable and hidden sectors with 
a $Z_2$--shift in a compactified coordinate, and is given by
$
g: (-1)^{(F_{1}+F_2)}\delta
$
~where the fermion numbers $F_{1,2}$ act on the spinorial
representations of the observable and hidden $SO(16)$ groups as
$
F_{1,2}:({\overline O}_{16}^{1,2},
             {\overline V}_{16}^{1,2},
             {\overline S}_{16}^{1,2},
             {\overline C}_{16}^{1,2})\longrightarrow~
            ({\overline O}_{16}^{1,2},
             {\overline V}_{16}^{1,2},
             -{\overline S}_{16}^{1,2},
             -{\overline C}_{16}^{1,2})
$
~and $\delta$ identifies points shifted by a $Z_2$ shift 
in the $X_9$ direction, {\it i.e.} 
$
\delta X_9 = X_9 +\pi R_9.~
$
The effect of the shift is to insert a factor of $(-1)^m$ into the lattice 
sum in eq. (\ref{zplus}), {\it i.e.} 
$
\delta:\Lambda_{m,n}^9\longrightarrow(-1)^m\Lambda_{m,n}^9.
$
~The second $Z_2$ acts as a twist on the internal coordinates 
given by 
$
{g^\prime}:(x_{4},x_{5},x_{6},x_7,x_8,x_9)
\longrightarrow
(-x_{4},-x_{5},-x_{6},-x_7,+x_8,+x_9). 
$
~The effect of the first $Z_2$ is to reduce the gauge symmetry from 
$E_8\times E_8$ to $SO(16)\times SO(16)$. 
The $Z_2^\prime$ twist reduces the number of 
spacetime supersymmetries from $N=4$ to $N=2$, 
and reduces the gauge symmetry arising from $SO(16)\times SO(16)$ 
to $SO(12)\times SO(4)\times SO(16)$. Additionally, it produces 
a twisted sector that gives rise to massless states in the spinorial ${\bf32}$ 
and ${\bf32}^\prime$, and vectorial ${\bf12}$, representations of $SO(12)$.  
In this vacuum the spinor--vector duality operates in terms of the
representations of $SO(12)\times SU(2)$ rather than in terms of 
representations of $SO(10)\times U(1)$, as the enhanced symmetry
point possess an $E_7$ symmetry rather than $E_6$. The spinor--vector
duality operates identically in the two cases and the case of the
single non--freely acting $Z_2$ twist elucidates more readily the
underlying structure of the spinor--vector duality.
The orbifold partition function is given by
$${Z~=~
\left({Z_+\over{Z_g\times Z_{g^{\prime}}}}\right)~=~
\left[{{(1+g)}\over2}{{(1+g^\prime)}\over2}\right]~Z_+}.$$
The partition function contains an untwisted sector and 
three twisted sectors. The winding modes in the sectors twisted by 
$g$ and $gg^\prime$ are shifted by $1/2$, and therefore these sectors only
produce massive states. The sector twisted by $g$
gives rise to the massless twisted matter states. 
The partition function has two modular orbits
and one discrete torsion $\epsilon=\pm1$. 
Massless states are obtained for vanishing lattice modes. 
The terms in the sector $g$ contributing to the massless 
spectrum take the form  
\beqn
& &     \Lambda_{p,q}
\left\{
 {1\over2}
\left( 
           \left\vert{{2\eta}\over\theta_4}\right\vert^4
         +
           \left\vert{{2\eta}\over\theta_3}\right\vert^4
\right)
\left[{
       P_\epsilon^+Q_s{\overline V}_{12}{\overline C}_4{\overline O}_{16}} +
   {P_\epsilon^-Q_s{\overline S}_{12}{\overline O}_4{\overline O}_{16} }
\right.
{\left.  \right] + }
\right. \nonumber\\ 
& &\nonumber\\ 
& &\left.
~~~~~~~~~~{1\over2}\left(    \left\vert{{2\eta}\over\theta_4}\right\vert^4
                      -
                         \left\vert{{2\eta}\over\theta_3}\right\vert^4\right)
\left[{
P_\epsilon^+Q_s
{\overline O}_{12}{\overline S}_4{\overline O}_{16}} \right.
{\left. \right] } 
\right\}~~~~~~~~~~~~~~~~~~~~~~~~~~~~
+~~\hbox{massive}
 \label{masslessterminpf}
\eeqn

where 
\beq
P_\epsilon^+~=~\left({{1+\epsilon(-1)^m}\over2}\right)\Lambda_{m,n}~~~;~~~
P_\epsilon^-=\left({{1-\epsilon(-1)^m}\over2}\right)\Lambda_{m,n} 
\label{pepluspeminus}
\eeq
Depending on the sign of the discrete torsion $\epsilon=\pm$ 
we note from  eq. (\ref{pepluspeminus}) that either the spinorial states, 
or the vectorial states, are massless. In the case with $\epsilon=+1$ 
we see from eq. (\ref{eplus}) that in this case massless momentum
modes from the shifted lattice arise in $P_\epsilon^+$ whereas 
$P_\epsilon^-$ produces only massive modes. Therefore, in his case  
the vectorial character ${\overline V}_{12}$ in eq. (\ref{pepluspeminus})
produces massless states, whereas the spinorial character
${\overline S}_{12}$ generates massive states.
In the case with $\epsilon=-1$ we note from eq. (\ref{eminus})
that exactly the opposite occurs. 
\beqn
{\epsilon~=~+1~~}&{\Rightarrow}&
{~~P^+_\epsilon~=~~~~~~~~~~~\Lambda_{2m,n}~~~
~~~~~~~~~~~~{ P^-_\epsilon~=~~~~~~~~
\Lambda_{2m+1,n}}~~~}\label{eplus}\\
{\epsilon~=~-1~~}&{\Rightarrow}&
{{ ~~P^+_\epsilon~=~~~~~~~~~~~\Lambda_{2m+1,n}}~~~
~~~~~~~~~P^-_\epsilon~=~~~~~~~~\Lambda_{2m,n}~~~}\label{eminus}
\eeqn
Another observation from the term appearing in eq. (\ref{masslessterminpf})
is the matching of the number of massless degrees of freedom in
the two cases. In the case with $\epsilon=-1$ the number
of degrees of freedom in the spinorial representation of
$SO(12)$ is 32. In the case with $\epsilon=+1$ the number of 
degrees of freedom in the vectorial representation of $SO(12)$ 
is 12. As seen from the first line in eq. (\ref{masslessterminpf})
the term in the partition function producing the vectorial states
also transforms as a spinor under the $SO(4)$ symmetry. Hence 
the total number of states is 24, {\it i.e.} there is still a mismatch
of 8 states between the two cases. However, we note from the 
second line in eq. (\ref{masslessterminpf}) that in the case
with $\epsilon=+1$ eight additional states are obtained from
the first excited states of the internal lattice. We note therefore
that the total number of degrees of freedom is preserved
under the duality map, {\it i.e.} 
$
{12\cdot 2+ 4\cdot2}{=}{32}
$

Given the relation of free fermionic models to toroidal orbifolds,
we can anticipate that the spinor--vector duality can be 
realised in terms of the moduli of the toroidal lattices. 
Those are the six dimensional metric, the antisymmetric 
tensor field and the Wilson lines \cite{narain}.
Indeed, the discrete torsion appearing in eq. 
(\ref{masslessterminpf}) can be translated to a
map between two Wilson lines \cite{ffmt}. We
note that in the case of (\ref{masslessterminpf}) the 
map between the Wilson lines is continuous. The reason
is the fact that we employed a single $Z_2$ twist 
on the internal coordinates. The moduli associated
with the Wilson line mapping are not projected out 
in this case and therefore the interpolation between
the two Wilson lines is continuous. In the more general 
case with a $Z_2\times Z_2$ twist these moduli are 
projected out and the mapping between the two Wilson
lines is discrete \cite{ffmt}. 

Additionally, we can understand
the spinor--vector duality in terms of a spectral flow operator
\cite{ffmt}, which may be generalised to
other cases. We recall that vacua with $E_6$
extended gauge symmetry are self--dual under the spinor--vector
duality, and that they correspond 
to vacua with $(2,2)$ worldsheet supersymmetry. Just like
the case of the worldsheet supersymmetry in the 
supersymmetric sector of the heterotic--string, there is a
spectral flow operator that acts as a generator of $E_6$ 
in the vacua with enhanced $E_6$ symmetry. On the supersymmetric
side the spectral flow operator mixes states with different
spacetime spin, whereas on the non--supersymmetric 
side it mixes states that differ by their $U(1)$ charge
in the decomposition $E_6\rightarrow SO(10)\times U(1)$, 
{\it i.e.} it mixes the states that transform as spinors
and vectors of $SO(10)$. When the $E_6$ symmetry is broken, 
{\it i.e.} when the worldsheet supersymmetry is broken from $(2,2)$ 
to $(2,0)$, the spectral flow operator induces the 
spinor--vector duality map between the two distinct vacua \cite{ffmt}. 

The spinor--vector duality is a novel symmetry that operates in the 
global space of $Z_2$ and $Z_2\times Z_2$ heterotic--string orbifolds
and provides valuable insight and interesting questions for future 
research. First, we note that the spinor--vector duality is 
a map between vacua that are completely unrelated in the 
effective field theory limit. For example, we may envision
a map between a model with 3 spinorial ${\bf16}$ representations,
and one vectorial ${\bf10}$ representation, to a model with 3 vectorial 
${\bf10}$ representations and one spinorial ${\bf16}$ representation. In terms 
of the low energy physics the two cases are fundamentally 
different. On the other hand, from the point of view of string theory
they are identical. Namely, there is an exact map from one to the 
other. The distinction between the string representation versus the 
effective field theory limit is that the string can access its 
massive modes, which are not seen in the effective field theory limit. 
Therefore, vacua that seem distinct in the effective field theory
limit are in fact related in the full string theory. We may 
further envision that at some early stage in the evolution
of the universe, when the heavy string modes are excited that the
two vacua can in fact mix. This possibility has implications on
the counting of distinct string vacua and therefore on the string
landscape. It is evident that our contemporary understanding
of the string landscape is still very rudimentary and we should 
proceed with caution before overstating our case.
The spinor--vector duality may also have interesting implications 
from a purely mathematical point of view. Namely, 
in the effective field theory limit there should exist a description
of the massless degrees of freedom in terms of a smooth effective
field theory {\it i.e.} in terms of a supergravity theory with a
classical geometry ({\it i.e.} some Calabi--Yau six dimensional 
manifold) with a vector bundle accounting for the gauge degrees
of freedom. The existence of the spinor--vector duality map
implies that there should be a similar map between the two 
effective theory limits of the two vacua. This is particularly
interesting in terms of the counting of the additional states
that are needed to compensate for the mismatch in the number 
of states between the two vacua. How do they arise
in the effective field theory limit? In the very least, 
the spinor--vector duality provides a valuable tool to 
study the moduli spaces of $(2,0)$ heterotic--string compactifications.

\section{other approaches}

The free fermionic models represents one of the approaches to 
string phenomenology. Several other approaches are being pursued, 
leading to overlapping and complementary results, in the
perturbative and nonperturbative domains. The literature 
on these subjects is vast and include several monographs, 
including, for example, \cite{ibanezuranga}.
A partial and incomplete
list of some of these studies include: 
geometrical studies \cite{gkmr, dopw, bmrw, hv, aglp};
orbifolds \cite{ino, blt, krz, lnrrrvw, bgrtv}; 
interacting CFTs \cite{ g, sy, gatos};
orientifolds \cite{csu, imr, kst}.
It should be emphasised that the present article
does not aim to review these important contributions, 
but merely those of the author. 
A comprehensive review is provided in reference \cite{ibanezuranga},
as well as in \cite{stringreviews}.

\section{Future} 

With the observation that the agent of electroweak symmetry 
breaking is compatible with an elementary scalar, 
particle physics and string phenomenology are set for a
bright future. In the particle physics realm the main 
questions are experimental. Are there additional
states associated with the electroweak symmetry 
breaking mechanism? {\it e.g.} Is spacetime 
supersymmetry realised in nature, and within reach of
contemporary colliders? Can we improve on the contemporary
measurements of the Standard Model parameters and by how
much? Can we build accelerators to probe 
energy scales in the deca--TeV region and above?
These are rather general questions and experiments 
should target more specific questions, {\it e.g.}
can we cool the muon phase space in a muon storage 
ring or a muon collider? The construction of a muon
based facility will advance the accelerator based technology
to a new era, and may be used as a Higgs factory in one
of its initial missions \cite{muonhiggs}. 

Particle physics and string phenomenology are two sides of the
same coin, and should not be regarded as distinct entities. 
Particle physics shows that experimental data can be parameterised 
by a model, which is based on the principles of point quantum field
theories, {\it i.e.} locality, causality and renormalizability.
This led to the development of the Standard Model, which is a 
quantum field theory with internal symmetries. A point quantum gravity
theory fails to satisfy these criteria. String theory resolves the 
problem with the third property by relaxing the first. 
String models provide consistent approaches to quantum
gravity, in which the internal symmetries are dictated 
by the consistency of the theory. As a common setting
for the gauge and gravitational interactions string theory
facilitates the calculation of the Standard Particle Model
parameters in a reduced framework. 

\subsection{Toward string predictions}

String theory leads to distinct signatures
beyond the Standard Model. In the first instance 
all the known stable string vacua at the Planck scale
are supersymmetric \cite{bachas}. Whether supersymmetry 
is manifested within reach of contemporary experiments
is a wild speculation. Nevertheless, this hypothesis 
is motivated on the ground that it facilitates extrapolation 
of the Standard Model parameters from the unification scale
to the electroweak scale. Furthermore, electroweak 
symmetry breaking at the low scale is generated 
in the supersymmetric scheme by the interplay of the 
top quark Yukawa coupling and the gauge coupling
of the strong interaction \cite{ewsb}.

Low scale supersymmetry is therefore not a necessary outcome
of string theory, but certainly its observation will provide 
further evidence that the different structures of string
constructions are realised in nature. Specific SUSY breaking 
scenarios in string models give rise to distinct supersymmetric
spectra and that in turn will be used to constrain 
further the phenomenological string vacua \cite{dedes}.
It is further noted that $R$--parity is generically broken 
in string vacua \cite{rparity} and that the LSP is not expected therefore
to provide a viable dark matter candidate. 

A generic prediction of string theory is the existence 
of additional gauge degrees of freedom, beyond those of the Standard 
Model, and is dictated by the consistency conditions
of string theory. However, construction of viable 
string models that allow for extra gauge symmetries
within reach of contemporary experiments is highly 
non--trivial. On the other hand, an extra $U(1)$ symmetry
may be instrumental to understand some phenomenological 
features of the Supersymmetric Standard Model, like the 
suppression of proton decay mediating operators and the
$\mu$--parameter. 

Another generic outcome of string models is the existence of 
exotic matter states. This feature of string constructions 
arises as a result of the breaking of the non--Abelian GUT symmetries
by Wilson lines, which 
results in exotic states that do not obey the 
quantisation rules of the original GUT group. Thus, one 
can get, for example, states that carry fractional electric 
charge. The lightest of the fractionally charged states is 
necessarily stable by electric charge conservations. The
experimental restrictions on states that carry fractional 
electric charge are severe and they must be either sufficiently
heavy, and/or sufficiently diluted to evade detection. 
Nevertheless, given that the bulk of the matter in the 
universe is dark, {\it i.e.} does not interact electromagnetically, 
stable string relics with a variety of properties can 
be contemplated \cite{ssr}. This includes for example the possibility that 
the string relics come as fractionally charged hadrons and
leptons, with charge $\pm1/2$. Such states will continue
to scatter in the early universe until they form a bound
hydrogen--like state with another fractionally charged companion. 
Provided that they are sufficiently heavy and sufficiently rare
they could have evaded detection by searches for rare isotopes.
Another possibility of exotic stable string relics arises 
when the $SO(10)$ GUT symmetry is broken to
$SU(3)\times SU(2)\times U(1)^2$. 
This case gives rise to states that carry the regular
Standard Model charges, but carry fractional charges
with respect to the extra $U(1)_{Z^\prime}\in SO(10)$.
This case, depending on the Higgs representations
that break the $U(1)_{Z^\prime}$, can result in discrete 
symmetries that forbid the decay of the exotic states
to the Standard Model states. It can therefore 
give rise to meta--stable heavy string relics that are
Standard Model singlets. Depending on the cosmological 
evolution in the early universe they could have been diluted
and reproduced as super--heavy states after reheating \cite{ssr}. 
Such states can produce viable dark matter \cite{ssr} candidates
as well as candidates for Ultra High Energy Cosmic Rays
(UHECR) \cite{cfp}. 

\subsection{Cosmological evolution} 

The early studies in string phenomenology, articulated
in section \ref{past}, entailed 
the in depth exploration of exemplary models and
the study of phenomenological properties. 
These studies focussed on the properties of the
massless spectra of these exemplary models
and led to the construction of the first known
Minimal Standard Heterotic String Models (MSHSM)
\cite{fny,cfn}.

The more recent studies, articulated in section
\ref{present}, involve the classification 
of large classes of models and the relations 
between them. The string vacua in this 
investigation are fermionic $Z_2\times Z_2$ 
orbifolds and are therefore related to
the exemplary models in section \ref{past}.
More importantly, the contemporary studies
involve the analysis of the partition 
functions associated with this 
class of string vacua. In that 
context they aim to explore how the 
massive string spectrum may play a role in 
the determination of the phenomenological 
and mathematical properties of string
models. This led to the discovery of 
spinor--vector duality in heterotic--string
models \cite{spinvecdual}.

One direction therefore in future string phenomenology studies
will involve the investigation of the associated string partition function, 
and in particular away from the free fermionic point. The 
most general form of the partition function affiliated 
with the $Z_2\times Z_2$ orbifolds, and hence
with the phenomenological free fermionic models 
is given by
\begin{eqnarray}
\non Z & = & \int
\frac{d^2\tau}{\tau_2^2}~\frac{\tau_2^{-1}}{\eta^{12}\bar\eta^{24}}~ 
\frac 1{2^3} 
\left(\sum { (-)^{a+b+ab}}
\Th{a}{b}\Th{a+{ h_1}}{b+{ g_1}}
         \Th{a+{ h_2}}{b+{ g_2}}
         \Th{a+{ h_3}}{b+{ g_3}}\right)_{\psi^\mu, \chi}\\
\nonumber &\times & 
\left(\frac 12 \sum_{\epsilon,\xi}
\Thbar{\epsilon}{\xi}^5
\Thbar{\epsilon+{ h_1}}{\xi+{ g_1}} 
\Thbar{\epsilon+{ h_2}}{\xi+{ g_2}}
\Thbar{\epsilon+{ h_3}}{\xi+{ g_3}}
\right)_{\bar\psi^{1\dots 5},\bar\eta^{{1},{2},{3}}}\\
\non & \times & \left(\frac 12 \sum_{H_1,G_1}\frac 12 \sum_{H_2,G_2}
(-)^{H_1G_1 + H_2G_2}
\Thbar{\epsilon+H_1}{\xi+G_1}^4
\Thbar{\epsilon+H_2}{\xi+G_2}^4\right)_{\bar\phi^{1\dots 8}}\\
\non & \times & \left(\sum_{s_i,t_i} \Gamma_{6,6}\left[^{h_i | s_i}_{g_i |
t_i}\right]\right)_{(y\omega \bar y\bar \omega)^{1\dots 6} } \times e^{i\pi
\Phi(\gamma,\delta,s_i,t_i,\epsilon,\xi,h_i,g_i,H_1,G_1,H_2,G_2)}
\nonumber
\end{eqnarray}
where the internal lattice is for one compact dimension is given 
by
\begin{equation}
 ~~~\Gamma_{1,1}[^h_g] ~=~ \frac R{\sqrt{\tau_2}} \sum_{\tilde m,n}
\exp\left[-\frac{\pi
R^2}{\tau_2}\left|\left(2\tilde m+
g\right)+\left(2n+h\right)\tau\right|^2\right], \nonumber
\end{equation}
and $\Phi$ is a modular invariant phase. The properties of the 
string vacua, away from the free fermionic point, can be explored 
by studying this partition function and the role of the 
massive states. Furthermore, while the current understanding 
of string theory is primarily limited to static solutions, 
exploration of dynamical scenarios can be pursued by 
compactifying the time coordinate on a circle and using 
the Scherk--Schwarz mechanism \cite{ss} in the compactified 
time--like coordinate. One then obtains a finite
temperature--like partition function that can be used 
to explore cosmological scenarios. Indeed, this is
the string cosmology program pursued by the Paris 
group over the past few years \cite{pariscos}.
In a similar spirit partition functions of 
string compactifications to two dimensions
have been explored revealing rich mathematical 
structures and the so--called massive supersymmetry, 
in which the massive spectrum exhibits Fermi--Bose 
degeneracy, whereas the massless spectrum does not
\cite{msds}. One can envision interpolations of
the two dimensional partition functions, associated 
with the cosmological and massive supersymmetry scenarios
to the four dimensional partition functions associated with the
phenomenological free fermionic models. The ultimate aim of this
program will be to explore possible mechanisms for dynamical
vacuum selection in string theory. 

\subsection{Dualities and fundamental principles}

Physics is first and foremost an experimental science. 
There is no absolute truth. There is only perception 
of reality as registered in an experimental 
apparatus\footnote{including astrophysical observations.}.
Be that as it may, the language that is used to interpret
the experimental signals is mathematics. 
The scientific methodology then entails:
\begin{itemize}
\item
the existence of some initial conditions,
which are either preset or set up in an experiment;
\item
the construction of a mathematical model 
that predict (or postdict) the outcome of the experiment;
\item
the confrontation of the predictions of the mathematical
model with the outcome of the experimental observations.
\end{itemize}
A successful mathematical model is the one that is able to account 
for a wider range of experimental observations. 
This scientific methodology has been developed over the past 
five hundred years or so. 

To construct a mathematical model one needs to define a 
set of variables that are to be measured experimentally.
The set of variables is key to the interpretation 
of the experimental outcome.
Over the years modern physics has undergone a process 
of evolution in terms of these basic set of variables.
In the Galilean--Newtonian system the basic set of variables
are the position and velocities. In modern
experiments the relevant measured variables are
typically the initial and final energy and momenta.
In the Lagrangian formalism the set of variable is
generalised to any set of configuration coordinates and
their derivatives with respect to time. In the 
Hamiltonian formalism the set of variables 
are the generalised configuration coordinates 
and their conjugate momenta, which constitute the 
phase space.  
This represents a nontrivial conceptual evolution from the 
Galilean--Newtonian system of position and velocities. 

String theory provides a consistent framework for
the perturbative unification of the gauge and gravitational
interactions. The string characterisation of  
the basic constituents of matter
reproduces the picture of elementary
particles with internal attributes.
String theory unifies the spacetime and internal
properties of elementary particles.
In the modern description of matter and interactions,
the three subatomic interactions are in a sense already 
unified. They are based on the gauge principle.
By giving rise to the mediators of the subatomic 
interactions that satisfy the gauge principle, 
and at the same time giving rise to the mediator 
of the gravitational interactions, that satisfy the 
gravitational gauge principle, string theory 
also unifies the principles underlying these theories. 

Can string theory be the final chapter in the unification
of the gauge and gravitational interactions. Unlike general 
relativity and quantum mechanics, string theory is not formulated
by starting from a fundamental principle and deriving
the physical consequences. Ultimately this is what we would like to
have. 

Perturbative and nonperturbative dualities have played 
a key role in trying to obtain a rigorous understanding 
of string theory. $T$--duality is an important perturbative 
property of string theory \cite{tduality}.
We may interpret $T$--duality and phase space duality 
in compact space. An additional important property of 
$T$--duality in string theory is the existence of 
self--dual states under $T$--duality. 

We may envision promoting phase--space duality
to a level of a fundamental principle.
This is the program that was undertaken in 
ref. \cite{fm2}. The key is the relation 
between the phase space variables via a generating 
function $S$, $p=\partial_q S$. To obtain a 
dual structure we define a dual generating 
function $T$, with $q=\partial_p T$.
The two generating functions are related by the 
dual Legendre transformations, 
\beq
S=p\partial_p T - T, 
\label{sdual}
\eeq
 and
\beq
T=q\partial_q S - S.
\label{tdual}
\eeq
Furthermore, one can show that $S(q)$ transforms as a
scalar function under the 
$GL(2,C)$--transformations
\beq
{\tilde q}={{A q + B}\over{ C q + D}}~,~~~
{\tilde p}= \rho^{-1}(C q + D)^2 p, 
\label{gl2ctrans}
\eeq
where $\rho=A D - B C\ne0$. We can associate the 
two Legendre transformations (\ref{sdual}) and 
(\ref{tdual}) with a second order differential 
equations whose solutions are $\{q\sqrt{p}; \sqrt{p}\}$
and $\{p\sqrt{q}; \sqrt{q}\}$, respectively.
A special class of solutions are those which 
satisfy the two sets of differential equations, {\it i.e.} 
$p=\gamma q$, with $\gamma={\rm constant}$. These are the 
self--dual solutions under the Legendre duality of 
(\ref{sdual}) and (\ref{tdual}).

Given that the Legendre transformations
are not defined for linear functions 
we have that the phase--space duality
is not consistent for physical systems with $S=A q + B$, 
{\it i.e.} precisely for the self--dual states.
It is further noted that the second order differential 
equations are covariant under coordinate transformations, 
but that their potential functions are only invariant 
under the M\"obius transformations (\ref{gl2ctrans}). 
This suggests the fundamental equivalence postulate 
\cite{fm2,fmijmp}:

{\it Given two physical systems labelled by
potential functions $W^a(q^a)\in H$ and 
$W^b(q^b)\in H$, where $H$ denotes the space
of all possible $W$'s, there always exists 
a coordinate transformations  $q^a\rightarrow q^b=v(q^a)$
such that $W^a(q^a)\rightarrow W^{av}(q^b)=W^b(q^b)$. }

This postulate implies that there should always exist 
a coordinate transformation connecting any state
to the state $W^0(q^0)=0$. Inversely, this means that 
any nontrivial state $W\in H$ can be obtained from the 
states $W^0(q^0)$ by a coordinate transformation. 

The classical Hamilton--Jacobi (HJ) formalism provides
a natural setting to apply this postulate. In the HJ formalism
a mechanical problem is solved by using canonical 
transformations to map the Hamiltonian 
of a nontrivial physical system, with nonvanishing kinetic
and potential energies, to a trivial Hamiltonian.
The solution is given by the Classical Hamilton--Jacobi 
Equation (CHJE) and the functional relation between the 
phase space variables is extracted by the relation 
$p=\partial_q S$, with $S$ being the solution of the 
HJ equation. We can pose a similar question, but imposing 
the functional relations $p=\partial_q S(q)$ on the trivialising
transformation $q\rightarrow q^0(q)$ and $S^0(q^0)=S(q)$.
This procedure is not consistent with the CHJE because 
the state $W^0(q^0)$ is a fixed point under the coordinate 
transformations \cite{fm2,fmijmp}. Consistency of the 
equivalence postulate therefore implies that the CHJE should
be deformed. Focussing on the stationary case, the most general 
deformation is given by
\beq
{1\over{2m}}\left({\partial_q S}_0\right)^2 + W(q) + Q(q) =0. 
\label{qshje}
\eeq
The equivalence postulate implies that eq. (\ref{qshje})
is covariant under general coordinate transformations. 
This is obtained provided that the combination $(W+Q)$ 
transforms as a quadratic differential. On the other hand 
all nontrivial states should be obtained from the state
$W^0(q^0)$ by a coordinate transformation. The basic 
transformation properties are then 
\beqn
W^v(q^v) & = & (\partial_{q^v}q^a)^2W^a(q^a)+(q^a;q^v), \nonumber\\
Q^v(q^v) & = & (\partial_{q^v}q^a)^2Q^a(q^a)-(q^a;q^v).\nonumber
\eeqn 
Comparing the transformations $q^a\rightarrow q^b \rightarrow q^c$
with $q^a\rightarrow q^c$ fixes the cocycle condition for the 
inhomogeneous term \cite{fmut},
\beq
(q^a;q^c)=(\partial_{q^c}q^b)^2
\left[ (q^a;q^b) - (q^c;q^b)\right]. 
\label{onedcocycle}
\eeq
The cocycle condition uniquely fixes the transformation properties of
the inhomogeneous term, and it is shown to be invariant 
under M\"obius transformations. In the one dimensional case
the M\"obius symmetry fixes the functional form of the 
inhomogeneous term to be given by the Schwarzian derivative 
$\{q^a,q^b\}$, where the Schwarzian derivative is given by
$$\{f(q),q\}= {f^{\prime\prime\prime}\over{f^\prime}}- {3\over2}\left(
{f^{\prime\prime}\over{f^\prime}}\right)^2.
$$
The Quantum Stationary Hamilton--Jacobi Equation (QSHJE) then takes
the form of a Schwarzian identity
\beq
\left(
\partial_q S_0
\right)^2= {\beta^2/ 2}
\left(
\left\{
\exp{(2iS_0/\beta)},q 
\right\}-
\left\{
S_0,q
\right\}
\right).
\label{schwarzid}
\eeq
With  $\beta=\hbar$ and the identifications
\beqn
W(q) = V(q) - E & = & -{\hbar^2\over{4m}}
\left\{
\exp{(2iS_0/\beta)},q 
\right\}, \\
Q(q) ~~~~~~~~~~~~~~~~~~~~~~ & = &
~~{\hbar^2\over{4m}}
\left\{
S_0,q
\right\}, 
\eeqn
the QSHJE takes the form 
\beq
{1\over{2m}} 
\left(
{{\partial S_0}\over {\partial q}}
\right)^2+V(q)-E +
{\hbar^2\over {4m}}
\left\{
S_0,q
\right\}=0, 
\eeq
which can be derived from the Schr\"odinger equation by
taking 
$$
\psi(q)= {1\over{\sqrt{S_0^\prime}}}{\rm e}^{\pm{{iS_0}\over\hbar}}.
$$
It is noted that the QSHJE is a non--linear differential equation,
whose solutions are given in terms of the two linearly independent
solutions, $\psi$ and $\psi_D$, of the corresponding Schr\"odinger
equation. Denoting $w=\psi_D/\psi$, and from the properties of the
Schwarzian derivative, it follows that the solution of the 
QSHJE is given, up to a M\"obius transformation, by
\beq
{\rm e}^{{{i2}\over\hbar}S_0\{\delta\}}=
{\rm e}^{i\alpha}
{{w+i{\bar\ell}}\over {w-i\ell}}
\eeq
where $\delta=\{\alpha,\ell\}$ with $\alpha\in R$ and ${\rm Re}\ell\ne0$, 
which is equivalent to the condition $S_0\ne const$. We note that the 
condition that the condition $S_0\ne const$ is synonymous with the 
condition for the definability of all phase space duality for 
all physical states. Thus, we find that the 
phase space duality and the equivalence postulate are intimately related. 
In essence, they are manifestation of the M\"obius symmetry 
that underlies quantum mechanics. It is further noted that the
trivialising map to the $W^0(q^0)$ state is
given by $q\rightarrow q^0=w$.

The equivalence postulate formalism reproduces the key
phenomenological properties of quantum mechanics, 
without assuming the probability interpretation of the 
wave function.
It implies that the momentum is real
also in the classically forbidden regions, hence 
implying the tunnelling effect of quantum mechanics
\cite{fmtqt, fmijmp}.
It implies quantisation of energy levels for bound
states with square integrable wave function. 
Additionally, it implies that time parameterisation  
of trajectories is ill defined in quantum mechanics \cite{fmtpqt}. 
The last two properties
are a direct consequence of the underlying M\"obius symmetry.
The M\"obius symmetry, which includes a symmetry under 
inversions, entails that space must be compact.
In the one dimensional case this is seen as imposing 
gluing conditions on the trivialising map at $\pm\infty$ 
\cite{fmtqt,fmijmp,fmmt}.
If space is compact the energy levels are always quantised.

The compactness of space also explains the inherent probabilistic 
nature of quantum mechanics, and the inconsistency of a 
fundamental trajectory parameterisation. 
There are two primary means to define time parameterisation of 
trajectories. In Bohmian mechanics \cite{holland,wyatt}
time parameterisation is obtained by identifying the 
conjugate momentum with the mechanical momentum,
{\it i.e.} $p=\partial_q S= m {\dot q}$, where $S$ is 
the solution of the quantum Hamilton--Jacobi equation.
In the classical Hamilton--Jacobi theory time parameterisation
is introduced by using Jacobi theorem
\beq
t={{\partial {\cal S}}_0^{\rm cl}\over{\partial E}}.
\label{jacobistheorem}
\eeq
In classical mechanics this is equivalent to identifying the conjugate
momentum with the mechanical momentum. Namely, setting
\beq
p=\partial_q{\cal S}_0^{\rm cl}=m{\dot q}
\eeq
yields
\beq
t-t_0=m\int_{q_0}^q{{dx}\over {\partial_x {\cal S}}_0^{\rm cl}}=
\int_{q_0}^q dx {{\partial~}\over{\partial E}}
{\partial_x {\cal S}}_0^{\rm cl}=
{{\partial {\cal S}_0^{\rm cl}}\over{\partial E}}.
\eeq
which provides a solution for the equation of motion $q=q(t)$.
Therefore, Bohmian mechanics brings back the notion of trajectories
for point particles.
%
%
However, the agreement between the definition 
of time via the mechanical time $p=m{\dot q}$, 
and its definition via Jacobi theorem (\ref{jacobistheorem})
is no longer valid in quantum mechanics. In quantum
mechanics we have 
\beq
t-t_0={{\partial S_0^{\rm qm}}\over{\partial E}} =
{{\partial~}\over{\partial E}} \int_{q_0}^q dx 
{\partial_x S}_0^{\rm qm}=
\left(m\over 2\right)
\int_{q_0}^q dx {{1-\partial_E Q}\over {\left(E-V-Q\right)^{1/2}}}.
\eeq
The mechanical momentum is then given by
\beq
m{{dq}\over {dt}}= m \left({{dt}\over{dq}}\right)^{-1}=
{{\partial_q S^{\rm qm}_0}\over{\left(1-\partial_E{\cal V}\right)}}
\ne\partial_qS_0^{\rm qm},
\eeq
where $\cal V$ denotes the combined potential 
${\cal V}=V(q)+Q(q)$. Therefore, in quantum mechanics the Bohmian time
definition does not coincide with its definition via Jacobi's theorem.

Floyd proposed to define time by using Jacobi's theorem
\cite{floyd}, {\it i.e.}
\beq 
t-t_0= {{\partial S_0^{\rm qm}}\over{\partial E}} .
\label{floydproposal}
\eeq 
Floyd's proposal would in principle provide a trajectory representation 
of quantum mechanics by inverting $t(q)\rightarrow q(t)$, which would
seem to be in contradiction with inherently probabilistic nature
of quantum mechanics.
However, if space is compact then the energy levels are 
always quantised, albeit with an experimentally indistinguishable
splittings \cite{de}. Hence, at a fundamental level
one cannot differentiate with respect to time, 
and time parameterisation of trajectories by Jacobi's theorem
is ill defined in quantum mechanics. Hence, time parameterisation 
of trajectories can only be regarded as an effective semi--classical
approximation and in that sense can provide a useful tool
in many practical problems \cite{wyatt}. 
The observation that time cannot be defined as a fundamental
variable in quantum mechanics may be extended to spacetime. 
That is the notion of spacetime may be a semi--classical approximate
notion rather than a fundamental one in quantum gravity. One should 
emphasise that statements such as ``time does not exist'' or 
``space is emergent'' are nonsensical. The physical question
is ``what are the relevant variables to parameterise the outcome
of experimental observations?''. Thus, the undefinability of time
parameterisation of trajectories in quantum mechanics
is at the heart of its probabilistic interpretation, which 
is well documented in experiments.

In this respect it is useful to provide an additional 
argument that shows that time parameterisation of trajectories 
is ill defined due to the M\"obius symmetry that underlies 
quantum mechanics, and consequently due to the compactness 
of space. In Bohmian mechanics the wave function is set as
\beq
\psi(q,t)=R(q){\rm e}^{iS/\hbar},
\label{psiq}
\eeq
where $R(q)$ and $S(q)$ are the two real functions of the 
QHJE, and $\psi(q)$ is a solution of the Schr\"odinger 
equation. The conjugate momentum is then given by 
$${\hbar}{\rm Im}{{\nabla\psi}\over\psi},$$
which we may use to define trajectories by
identifying it with $m{\dot q}$. The flaw 
in this argument is in the Bohmian identification of 
the wave function by (\ref{psiq}). The issue is 
precisely the boundary conditions imposed by
the M\"obius symmetry that underlies quantum 
mechanics and the compactness of space. If space 
is compact then the wave function is necessarily a 
linear combination of the two solutions of the 
Schr\"odinger equation
\beq
\psi= R(q) \left( A {\rm e}^{{i\over \hbar} S} + 
B {\rm e}^{-{i\over \hbar} S}\right),
\label{onedwavefunction}
\eeq
albeit one of the coefficients $A$ or $B$ can be very small,
but neither can be set identically to zero. In this case
$$\nabla S\ne {\hbar}{\rm Im}{{\nabla\psi}\over\psi}$$
and the Bohmian definition of trajectories is invalid.

The equivalence postulate approach
therefore reproduces the main phenomenological characteristics
of quantum mechanics. In fact, in retrospect this is not a
surprise. It may be regarded as conventional quantum 
mechanics with the addendum that space is compact, 
as dictated by the M\"obius symmetry that underlies the 
formalism. 

The one dimensional case reveals the M\"obius symmetry that
underlies the equivalence postulate and hence underlies quantum 
mechanics.
The equivalence postulate formalism extends to the higher dimensional case
both with respect to the Euclidean and Minkowski metrics \cite{bfm}.
For brevity I summarise here only the non--relativistic case. 
The relativistic extensions as well as the generalisation
to the case with gauge coupling are found in ref. \cite{bfm}. The key to 
these extensions are the generalisations of the cocycle condition eq. 
(\ref{onedcocycle}), and of the Schwarzian identity eq. (\ref{schwarzid}). 
Denoting the transformations between two sets of coordinate systems 
by
\beq
q\rightarrow q^v = v(q)
\label{qtovq}
\eeq
and the conjugate momenta by the generating function $S_0(q)$,
\beq
p_k= {{\partial S_0}\over {\partial q_k}}. 
\label{pkmomenta}
\eeq
Under the transformations (\ref{qtovq}) we have $S_0^v(q^v)=S_0(q)$, hence
\beq
p_k\rightarrow p_k^v=\sum_{i=1}^D J_{ki}p_i
\label{pkv}
\eeq
where  $J$ is the Jacobian matrix
\beq
J_{ki}={{\partial q_i}\over {\partial q_j^v}}.
\label{jacobian}
\eeq
Introducing the notation
\beq
(p^v|p)={{\sum_k (p_k^v)^2}\over{\sum_kp_k^2}}={{p^tJ^tJp}\over {p^tp}}. 
\label{pvp}
\eeq
 the cocycle condition takes the form 
\beq
(q^a;q^c)=(p^c|p^b)\left[(q^a;q^b)-(q^c;q^b)\right], 
\label{cocycleinEspace}
\eeq
which captures the symmetries that underly quantum mechanics. 
It is shown that the cocycle condition, eq. (\ref{cocycleinEspace})
is invariant under $D$--dimensional M\"obius transformations, 
which include dilatations, rotations, translations and reflections in
the unit sphere \cite{bfm}. The quadratic identity, eq.
(\ref{schwarzid}), 
is generalised by the basic identity
\beq 
\alpha^2(\nabla S_0)^2=
{\Delta(R{\rm e}^{\alpha S_0})\over R{\rm e}^{\alpha S_0}}-
{\Delta R\over R}-{\alpha\over R^2}\nabla\cdot(R^2\nabla S_0), 
\label{ddidentity}
\eeq
which holds for any constant $\alpha$ and any functions 
$R$ and $S_0$. Then, if $R$ 
satisfies the continuity equation 
\beq
\nabla\cdot(R^2\nabla S_0)=0, 
\label{conteq}
\eeq
and setting $\alpha=i/\hbar$, we have 
\beq
{1\over2m}(\nabla S_0)^2=-{\hbar^2\over2m}{\Delta(R{\rm e}^{{i\over 
\hbar} S_0})\over R{\rm e}^{{i\over\hbar}S_0}}+
{\hbar^2\over2m}{\Delta R\over R}. 
\label{identity2}
\eeq 
In complete analogy with the one dimensional case we make identifications, 
\beqn
W(q)=V(q)-E& = &{\hbar^2\over2m}{\Delta(R{\rm e}^{{i\over\hbar}S_0})\over 
R{\rm e}^{{i\over \hbar}S_0}}, 
\label{ddwq}\\
Q(q)& =& -{\hbar^2\over2m}{\Delta R\over R}. 
\label{identity3}
\eeqn
Eq. (\ref{ddwq}) implies the $D$--dimensional Schr\"odinger equation
\beq
\left[-{\hbar^2\over2m}{\Delta}+V(q)\right]\Psi=E\Psi. 
\label{ddschroedingereq}
\eeq
and the general solution
\beq
\Psi= R(q) \left( A {\rm e}^{{i\over \hbar} S_0} + 
B {\rm e}^{-{i\over \hbar} S_0}\right).
\label{ddwavefunction}
\eeq
is mandated by consistency of the equivalence postulate. 
We note that the key to these generalisations is
the symmetry structure that underlies the formalism.
Seeking further generalisation of this approach simply entails that 
this robust symmetry structure is retained.

\subsubsection{The classical limit} 

The invariance of the cocycle condition under
M\"obius transformations may only be implemented  if space is compact. 
The decompactification limit may 
represent the case when the spectrum of the free quantum particle becomes
continuous. In that case time parameterisation of quantum trajectories
is consistent with Jacobi's theorem \cite{floyd,fmijmp,fmtpqt}. 
However, the decompactification 
limit can be seen to coincide with the classical limit. For this purpose
we examine again the case of the free particle in one dimension.
This is sufficient since all physical states can be mapped to
this state by a coordinate transformation. 
The quantum potential associated with the state
$W^0(q^0)\equiv0$ is given by
\beq
{Q}^0={\hbar^2\over4m}\{S_0^0,q^0\}= 
-{{\hbar^2 ({\rm Re}\,\ell_0)^2}\over2m}
{1\over{\vert q^0-i\ell_0\vert^4}}.
\label{Qnotpot}
\eeq
It is noted that the limit $q^0\rightarrow\infty$ coincides with the 
limit ${Q}^0\rightarrow 0$, {\it i.e.} with the classical limit \cite{tqc}.
This observation is consistent with the recent claim that the 
universe cannot be closed classically \cite{aharon}.
Possible signatures for nontrivial topology in the CMB has been 
of recent interest \cite{cmb}. Further experimental support for the
equivalence postulate approach to quantum mechanics may arise from 
modifications of the relativistic--energy momentum 
relation \cite{opera}, which affects the 
propagation of cosmic gamma rays \cite{aemns}.

\subsubsection{Where is the connection with string theory?} 

The simple answer to the question may be: in the future. 
Nevertheless, we may attempt to gather some hints how the 
connection may exist. 
String theory is a self--consistent perturbative framework 
for quantum gravity. As such it provides an effective
approach, but is not formulated from a fundamental principle. 
An important property of string theory is T--duality, 
which may be interpreted as phase space duality in compact 
space. We may conjecture that phase space duality is the 
fundamental principle and use that as a starting point for 
formulating quantum gravity. This is what the equivalence 
postulate approach aims at.

Consistency of the equivalence postulate approach 
dictates that the CHJE is replaced by the QHJE 
and that the quantum potential $Q(q)$ is never zero. 
In the one dimensional case the 
Schwarzian derivative may be interpreted as a curvature 
term \cite{flanders, fmqt,de}. In the higher dimensional case
it is proportional to the curvature of the function $R(q)$. 
Thus, we may interpret the quantum potential as
an intrinsic curvature term associated with an elementary
particle. Point particles do not have curvature. Hence, 
the interpretation of the quantum potential as a curvature
term hints to the connection with internal structure
of elementary particles.

The M\"obius symmetry underlying quantum mechanics 
in the equivalence postulate formalism also implies
the existence of a finite length scale \cite{fmpl}. 
For this purpose we can again 
study the one dimensional stationary case
with ${W}^0(q^0)=0$. 
The Schr\"odinger equation takes the form
$${\partial^2\over{\partial q}^2}\psi=0.$$
The two linearly independent solutions are $\psi^D=q^0$ and 
$\psi=const$. Consistency of the equivalence postulate 
mandates that both solutions must be retained.
The solution of the corresponding QHJE is given by \cite{fm2,fmijmp}  
$$
{\rm e}^{{2i\over\hbar}S_0^0}={\rm e}^{i\alpha}
{{q^0+i{\bar\ell}_0}\over{q^0-i\ell_0}}, 
$$
where $\ell_0$ is a constant with the dimension of length 
\cite{fmpl,fmijmp}, 
and the conjugate momentum 
$p_0=\partial_{q^0} S_0^0$ 
takes the form
\beq
p_0=\pm{{\hbar (\ell_0+{\bar\ell}_0) }\over {2\vert q^0- i\ell_0\vert^2}}.
\label{pzerozero}
\eeq
It is noted that $p_0$ vanishes only for $q^0\rightarrow\pm \infty$.
The requirement that in the classical limit
$\lim_{\hbar\rightarrow0}p_0=0$ suggests
that we can set \cite{fmpl,fmijmp}
\beq
{\rm Re}\,\ell_0=\lambda_p= \sqrt{{\hbar G}\over c^3}, 
\label{setell0}
\eeq
{\it i.e.} we identify ${\rm Re}\,\ell_0$ with the Planck length.
The invariance under the M\"obius transformations
mandates the existence of a finite length scale.
Additionally, from eq. (\ref{pzerozero}) follows that $p_0$ is maximal
for $q^0=-{\rm Im}\ell_0$, {\it i.e.}
\beq
\vert p_0(-{\rm Im}\ell_0)\vert ={\hbar\over{{\rm Re}\ell_0}}.
\label{pzeromax}
\eeq
The equivalence postulate mandates that ${\rm Re}\ell_0\ne 0$. 
Consequently, $p_0$ is always finite and $\ell_0$ acts as an
ultraviolet cutoff. As we would expect, the existence
of an ultraviolet cutoff is tightly linked to the 
existence of a finite length scale. The fundamental 
feature is the M\"obius symmetry at the core of 
the quantum mechanics.

%


\section{Conclusions}

The indication from the LHC of a scalar resonance compatible
with perturbative electroweak symmetry breaking reinforces the
Standard Model parameterisation of all subatomic experimental 
data. The logarithmic evolution of the Standard Model gauge and
matter parameters suggests that the Standard Model provides 
a viable parameterisation up to the Planck scale. Supersymmetry
preserves the logarithmic running also in the scalar sector, 
which provides reasonable motivation to seek experimental 
evidence for its validity in the LHC, VLHC (Very Large Hadron Collider)
and other future 
machines. It should be stressed that the viability of the
experimental program rests on its ability to deliver 
a working machine in the first place and to measure 
the parameters of the Standard Model to better accuracy
in the second.
Discovering new physics is an added bonus. 

The Planck scale is an ultraviolet cutoff, at which 
gravitational effects are of comparable strength to the 
gauge interactions. String theory provides a perturbatively
consistent framework that incorporates gravity and the 
gauge interactions and enables the construction of 
phenomenological models. The state of the art in this 
regard are string models that reproduce the spectrum
of the Minimal Supersymmetric Standard Model. 
The understanding of string theory as well as 
that of the space of string solutions is still 
at its infancy. As long as the experimental data
does not indicate that this is on the wrong track,
its exploration continues to be of interest.
Ultimately, in the future we would like to formulate 
quantum gravity from a fundamental principle. Phase space
duality and the equivalence postulate of quantum mechanics
provide a good starting point for that purpose. 


\acknowledgements{Acknowledgements}

I would like to thank
Jorge Lopez, Dimitri Nanopoulos, Kajia Yuan, Edi Halyo,
Ben Grinstein, Philip Argyres, Keith Dienes, John March--Russell,
Claudio Coriano,
Sangyeon Chang, Marco Matone, Jogesh Pati, John Ellis,
Gerald Cleaver, Per Berglund, Zongan Qiu, Maxim Pospelov,
Richard Garavuso, Jose Isidro, Alessandro Cafarella, Sander Nooij,
Carlo Angelantonj, Ron Donagi,
John Rizos, Costas Kounnas, Cristina Timirgaziu, Elisa Manno, Marco Guzzi,
Ben Assel, Kyriakos Christodoulides, Mirian Tsulaia, Thomas Mohaupt, Radu Tatar,
William Walters, Viraf Mehta, Hassan Sonmez, Laura Bernard, Ivan Glasser,
Panos Athanasopoulos, Doron Gepner
and many others for collaborations and discussions.
This work is supported in part by the STFC under contract ST/J000493/1.


\conflictofinterests{Conflicts of Interest}

The author declares no conflicts of interest. 

\bibliographystyle{mdpi}
\makeatletter
\renewcommand\@biblabel[1]{#1. }
\makeatother


\end{document}